\documentclass[aps,twocolumn,amsmath,amssymb,preprintnumbers]{revtex4}
\usepackage{amsmath} \usepackage{amsfonts} \usepackage{amssymb}
\usepackage{bbm}
\usepackage{epsfig}
\usepackage{graphics}
\usepackage{graphicx}
\newcommand{\be}{\begin{equation}}
\newcommand{\ee}{\end{equation}}
\newcommand{\ba}{\begin{eqnarray}}
\newcommand{\ea}{\end{eqnarray}}
\newcommand{\nn}{\nonumber}
\newcommand{\kr}{\rangle}
\newcommand{\kl}{\langle}

\newcommand{\tr}{\textup{tr}}

\newcommand{\td}{\tilde\Delta}

\begin{document}

\title[ ]{Quantum particles from coarse grained classical probabilities in phase space}

\author{C. Wetterich}
\affiliation{Institut  f\"ur Theoretische Physik\\
Universit\"at Heidelberg\\
Philosophenweg 16, D-69120 Heidelberg}

\begin{abstract}
Quantum particles can be obtained from a classical probability distribution in phase space by a suitable coarse graining, whereby simultaneous classical information about position and momentum can be lost. For a suitable time evolution of the classical probabilities and choice of observables all features of a quantum particle in a potential follow from classical statistics. This includes interference,  tunneling and the uncertainty relation. 
\end{abstract}

\maketitle

\section{Introduction}
\label{Intro}

Coarse graining is a fundamental method for describing situations where the complete information of a system is not relevant or available for a physical problem. One example are block spins 
\cite{K} or functional renormalization \cite{Wi}, \cite{CWAAA}, where one averages over microscopic degrees of freedom in order to describe macroscopic physics. Another example are observables  which only depend on the degrees of freedom of a local subsystem, which is embedded in a larger environment. We use here the concept of ``coarse graining'' in a generalized sense in order  to {\it define} a subsystem with less effective degrees of freedom than the total system. The physics of the subsystem is described by effective laws which do not refer any longer to the total system. Often these effective laws are qualitatively different from the laws for the total system from which they are derived. A well known example is the entropy in quantum mechanics. It is conserved for an isolated total system, while it may vary in time for a subsystem.

In this paper we describe quantum mechanics  within the basic setting of a classical statistical ensemble, as characterized by a positive semi-definite
probability distribution. As for a classical particle the states of the total system correspond to points in phase space. For a particular class of subsystems of this ensemble we will find qualitatively new laws which actually describe quantum physics. A typical feature of subsystems are observables whose expectation values can be computed from the information available for the subsystem, while this does not hold for their classical correlation functions. In this case the subsystem is described by ``incomplete statistics'' \cite{3}. Sequences of measurements of such observables have to be described by new ``measurement correlations'' which are computable from the information available for the subsystem \cite{CWAA,CW2,CWE}. The measurement correlations differ from the classical correlations and can violate Bell's inequalities \cite{Bell}, \cite{BS}. The coarse graining to a subsystem constitutes the basis for many of the surprising properties of quantum physics. Indeed, the conceptual issues how quantum systems can be consistently derived as subsystems of classical statistical ensembles have already been clarified \cite{CWAA,CW2,CWE}.

In the present paper we consider the probabilistic description of one particle in a potential. The classical ensemble for the total system is determined by the probability distribution in phase space, $w(x,p)\geq 0$, normalized according to $\int_{x,p}w(x,p)=1$. (Here $x$ and $p$ are the three-dimensional position and momentum variables and $\int_p$ contains the appropriate factor $(2\pi)^{-3}$, with units $\hbar=1$.) We will discuss a coarse graining to a subsystem where the simultaneous availability of information on both position and momentum is lost (except for particular circumstances). In general, neither the classical position nor the classical momentum observables are computable from the information of the subsystem. We will describe new observables for position and momentum which remain computable for the subsystem. These will turn out to be the position and momentum observables of a quantum particle. We also formulate a fundamental equation for the time evolution of the classical probability density $w(x,p)$ which is consistent with the coarse graining. It  results in the Heisenberg equation for the time evolution of the expectation values of quantum position and momentum and their correlations. All features of a quantum particle in a potential, including interference, tunneling and uncertainty relation, follow in this way from a classical statistical setting.

An important ingredient in our formulation is the ``classical wave function'' $\psi_C(x,p)$ in phase space \cite{3A}. This real function is  related to the probability density by $w(x,p)=\psi^2_C(x,p)$. 
In principle, the sign of $\psi$ is free, but continuity requirements severely restrict the relative sign between neighboring points in phase space \cite{3A}. If the remaining sign ambiguity has no physical relevance, we may consider $w(x,p)$ or $\psi_C(x,p)$  as equivalent descriptions of the probability distribution of the ensemble. The classical wave function  allows us to define a ``quantum transform'' of the probability density in phase space
\be\label{AA1}
\bar\rho_w(x,p)=\int
\psi_C(x+\frac r2,p+s)\psi_C(x+\frac{r'}{2},p+s')\cos (s'r-sr'),
\ee
where the integral extends over position variables $r,r'$ as well as momentum variables $s,s'$. We will find that the quantum transform $\bar\rho_w$ has the properties of the Wigner-transform of the density matrix for a quantum particle, from which the expectation values of quantum operators can be computed in a standard way.

If the time evolution of the classical probability distribution obeys the differential evolution equation
\ba\label{AA2}
\partial_t\psi_C(x,p)&=&-\frac pm\partial_x\psi_C(x,p)+K(x,\partial_p)\psi_C(x,p),\\
K&=&-i\left[V\left(x+\frac i2\partial_p\right)-V\left(x-\frac i2\partial_p\right)\right],\nn
\ea
one can infer that the time dependence of $\bar\rho_w$ follows the standard unitary time evolution for a quantum particle in a potential $V$. For harmonic potentials eq. \eqref{AA2} describes the classical evolution according to the Liouville equation, while in the more general case $K$ is a real operator which is no longer linear in the momentum derivative $\partial_p$. Postulating 
eq. \eqref{AA2} as the fundamental law for the time evolution of the probability density in phase space, all features of a quantum particle in a potential follow. In particular, the concepts of classical trajectories and Newton's laws follow only in the ``classical limit'' when the action is large as compared to $\hbar=1$.

Our implementation of a quantum particle within classical statistics is not a deterministic local hidden variable theory but rather assumes that the fundamental description of the real world is intrinsically probabilistic \cite{GenStat}.
In ref. \cite{CWAA,CW2,CWE} the basic conceptual and formal settings have been described in detail for statistical ensembles which correspond to two-state and four-state quantum mechanics. It was shown how the quantum formalism with non-commuting operators arises from a description of a subsystem. We have discussed explicitly quantum entangled states \cite{CW1}, \cite{CWAA}, \cite{CWE} and shown that the measurement correlation  is equivalent to the usual quantum correlation. It may  violate Bell's inequalities \cite{Bell}. In this context the property of incomplete statistics \cite{3}, where the measurement correlation is not based on joint probabilities, is crucial \cite{CW2}, \cite{CWAA}. (Complete statistical systems, for which the measurement correlation employs the joint probabilities, have to obey Bell's inequalities \cite{BS}.)  The use of probabilistic observables \cite{PO} for the subsystem avoids conflicts with the Kochen-Specker theorem \cite{KS}, as demonstrated explicitly in \cite{CWAA}.

The difference between a quantum particle and a classical particle reflects the issue of incompleteness or completeness of the statistical description. For a quantum particle the joint probabilities for position and momentum
are not available. This leads to non-commuting operators describing the position and the momentum of a particle and to Heisenberg's uncertainty relation. In contrast, for  a classical particle
the joint probabilities for position and momentum are available.
 Now the position and momentum operators commute and a simultaneous sharp measurement for both types of observables becomes possible. For both the quantum particle and the classical particle the formalism of quantum mechanics can be used. The description of the classical particle employs, however, an unusual set of observables where location and momentum operators commute. 

The present paper is the third in a series of papers devoted to the description of quantum particles in a classical statistical setting. The first paper \cite{3A} addresses the issue on a rather fundamental level by discussing how yes/no questions for the position and motion of a particle can be asked. Already at this level it becomes apparent that both classical and quantum particles can be described within the same conceptual framework. In particular, one may employ the quantum formalism based on a wave function or density matrix for both types of particles. Only the relevant observables and the time evolution of the probability density differ. In a second paper \cite{CWQP2} we discuss explicitly the time evolution of the probability density in phase space and the choice of observables which lead to a description of a quantum particle in terms of a classical statistical ensemble.
Since there are no conceptual differences between classical and quantum particles we can introduce zwitters as particles that are neither purely quantum nor purely classical. As a function of a ``zwitter parameter'' $\sin^2\gamma$ a continuous interpolation between a quantum particle for $\gamma=0$ and a classical particle for $\gamma=\pi/2$ becomes possible.  It will be an experimental challenge to put bounds on $\gamma$. A short account of our findings is given in \cite{CWZ}.

The present paper motivates our proposal for the evolution equation and the choice of position and momentum observables by a procedure of coarse graining. It is organized as follows. We summarize in sections II and III  the properties of the classical wave function and statistical observables \cite{3A} that will be needed in the following. Sect. IV discusses the coarse graining of the classical probability distribution which results in a quantum density matrix $\rho_Q$ for the subsystem. We discuss the position and momentum observables that can be defined on the coarse grained level. This means that their expectation values can be computed from the density matrix $\rho_Q$ alone, without invoking information from $w(x,p)$ which is no longer available on the level of the subsystem. We find that these observables do not commute and have all the properties of the position and momentum observables for a quantum particle. In particular, they obey Heisenberg's uncertainty relation.

At this level the time evolution of the classical probability distribution in phase space is not yet specified. In sect. V we start with the time evolution appropriate for a classical particle, as given by the Liouville equation. In general, the time evolution of $\rho_Q$ cannot be described in this case by a differential equation that only involves $\rho_Q$. The time evolution of the subsystem does therefore not decouple from the ``total system''. In particular, the time evolution of $\rho_Q$ is, in general, not unitary. We therefore extend our discussion to modified evolution equations for the probability density $w(x,p)$, describing ``generalized particles''. In sect. \ref{Quantum particle from classical probabilities} we show that the particular choice of the evolution equation \eqref{AA2} leads to a unitary evolution of $\rho_Q$, with Hamiltonian given by the usual quantum Hamiltonian for a particle in a potential. We establish that a classical probability density obeying   eq.~\eqref{AA2} admits a coarse graining for which the time evolution of the subsystem can be described in terms of the information available for the subsystem alone. If one also uses the position and momentum observables which are appropriate for the subsystem, all laws of quantum mechanics for a particle in a potential follow.

In sect.~VII we turn to the question under which circumstances the classical position and momentum observables may remain computable from the subsystem alone. Perhaps surprisingly, we find that  the classical  probability density $w(x,p)$ can be uniquely reconstructed if the density matrix $\rho_Q$ for the subsystem corresponds to a pure quantum state. In this case the classical observables remain computable, as well as interpolating ``sharpened observables'' that we discuss in sect. VIII. Sect.~IX is devoted to our conclusions.

\section{Classical wave function}
\label{classicalparticle}

Our starting point is the probability density for one particle in phase space, $w(z,p)$, which is positive everywhere and normalized
\be\label{2A}
w(z,p)\geq 0~,~\int_{z,p}w(z,p)=1.
\ee
(We use from now on $z$ for the position coordinate in phase space in order to distinguish it from the argument of the quantum wave function  $\psi_Q(x)$ in position space.) Expectation values of classical observables obey the usual rule
\be\label{2B}
\langle F(z,p)\rangle=\int_{z,p}F(z,p)w(z,p).
\ee
We define the real wave function in phase space, $\psi_C(z,p)$, as the positive or negative root of $w(z,p)$,
\be\label{2C}
w(z,p)=\psi^2_C(z,p).
\ee
The choice of sign of $\psi(z,p)$ is strongly constrained by continuity properties, such that up to a trivial overall sign the wave function $\psi_C$ carries the same information as $w(z,p)$ \cite{3A}.

In terms of the phase space wave function $\psi_C$ we can write eq.~\eqref{2B} in the form usual for the quantum formalism
\be\label{2D}
\langle F(z,p)\rangle=\int_{z,p}\psi_C(z,p)F(\hat X_{cl},\hat P_{cl})\psi_C(z,p).
\ee
Here we have introduced the commuting classical position and momentum operators $\hat X_{cl}$ and $\hat P_{cl}$. The wave function $\psi_C(z,p)$ is a simultaneous eigenstate to $\hat X_{cl}$ and $\hat P_{cl}$
\ba\label{zA}
\hat X_{cl}\psi_C(z,p,t)&=&z\psi_C(z,p,t),\nn\\
\hat P_{cl}\psi_C(z,p,t)&=&p\psi_C(z,p,t).
\ea

The introduction of the ``classical wave function'' $\psi_C(z,p)$ allows us to describe a classical particle in the quantum formalism  \cite{3A}. Indeed, the structure of quantum mechanics perfectly allows for commuting position and momentum operators, which are represented here as
\be\label{zB}
\hat X_{cl}=z\delta(z-z')~,~\hat P_{cl}=p\delta(p-p').
\ee
(In our normalization $\delta(p-q)$ stands in $d$ dimensions for $(2\pi)^d\delta^d(p-q)$ corresponding to $\int_p=(2\pi)^{-d}\int d^dp$.) We can also write the time evolution of the probability density $w(z,p)$ in the form of a Schr\"odinger equation for the classical wave function. For a classical particle the probability density evolves according to the Liouville equation
\ba\label{H5}
&&\partial_tw(z,p)=-\hat L w(z,p),\nn\\
&&\hat L=\frac pm\partial_z-\frac{\partial V}{\partial z}\partial_p.
\ea
Since $\hat L$ is linear in the derivatives, this extends to
\be\label{H6}
\partial_t\psi_C(z,p)=-\hat L\psi_{C}(z,p).
\ee
The Schr\"odinger equation in phase space obtains by multiplication with $i\hbar$,
\be\label{H7}
i\hbar\partial_t\psi_C=H_L\psi_C,
\ee
with
\be\label{H8}
H_L=-i\hbar\hat L=-i\hbar\frac pm\partial_z+i\hbar
\frac{\partial V}{\partial z}\partial_p
\ee
The Hamiltonian $H_L$ is a hermitean operator as it should be. 

The choice of the Hamiltonian $H_L$ leads to the commutation relations
\be\label{zD}
[H_L,\hat X_{cl}]=-\frac{i\hbar}{m}\hat P_{cl}~,~[H_L,\hat P_{cl}]=i\hbar
\frac{\partial V}{\partial\hat X_{cl}}.
\ee
These are the standard equations for the time evolution of observables in the Heisenberg picture of quantum mechanics. (We recall that $H_L$ denotes here the Hamilton operator within the quantum formalism for a classical particle and should not be confounded with the classical Hamiltonian in classical mechanics.) A standard quantum mechanical calculation yields then the time evolution for the expectation values 
$\bar x_{cl}=\kl \hat X_{cl}\kr$ and $\bar p_{cl}=\kl\hat P_{cl}\kr$, according to 
\be\label{zE}
\partial_t\bar x_{cl}=\frac{\bar p_{cl}}{m}~,~\partial_t\bar p_{cl}=-\kl\frac{\partial V}{\partial z}\kr.
\ee
These are the same evolution equations for the expectation values of position and momentum as for a quantum particle. They reduce to classical trajectories if $\kl\partial V/\partial z\kr$ can be replaced by $(\partial V/\partial z)(\bar x_{cl})$. We emphasize that the second eq. \eqref{zE} is the classical statistical equation which  describes the time evolution for a distribution of initial conditions. 

The difference between a classical and a quantum particle does not arise from the different formal structure between quantum mechanics and classical statistics. We have  established  how to describe a classical particle within the formalism of quantum mechanics. The difference between quantum or classical behavior of a particle is rather due to the different dynamics or, in other words, to different Hamiltonians, and to the use of different types of operators.

\section{Statistical observables}

The concept of the wave function in phase space allows the introduction of derivative operators as familiar from quantum mechanics. We may
 consider the hermitean operators
\be\label{113A}
\hat P_s=-i\partial_z~,~\hat X_s=i\partial_p
\ee
which obey
\ba\label{12A}
[\hat P_s,\hat X_{cl}]&=&-i~,~[\hat X_{s}~,~\hat P_{cl}]=i~,~[\hat X_{s},\hat P_s]=0,\nn\\
~[\hat P_s,\hat P_{cl}]&=&0~,~[\hat X_s,\hat X_{cl}]=0,
\ea
and
\be\label{113B}
H_L=\frac1m\hat P_{cl}\hat P_s+V'(\hat X_{cl})\hat X_s.
\ee
Their expectation values are computed according to the usual rule,
\be\label{13A}
\kl F(\hat X_s,\hat P_s)\kr=\int_{z,p}\psi_C(z,p)F(\hat X_s,\hat P_s)\psi_C(z,p),
\ee
and vanish by the absence of boundary terms
\ba\label{113C}
\kl  P_s\kr=0~,~\kl  X_s\kr=0.
\ea
Nevertheless, the squared operators have positive, in general nonzero, expectation values
\ba\label{113D}
\kl P^2_s\kr&=&\frac14\int_{x,p}w(\partial_z\ln w)^2,\nn\\
\kl X^2_s\kr&=&\frac14
\int_{x,p}w(\partial_p\ln w)^2.
\ea

Expressed in terms of the probability density $w$ the operators  $\hat X_s,~\hat P_s$ do not correspond to classical observables in the usual sense. Their expectation value is not given by an expression linear in $w$, as in eq.~\eqref{2B}. We can formally write
\be\label{QC26}
\langle P^2_s\rangle=\frac14\kl \tilde\Delta^2\kr~,~\tilde\Delta=\partial_z\ln w,
\ee
but the ``observable'' $\tilde\Delta$ is now a ``statistical observable'' which depends on $w$. The status of  $\td^2$ is somewhat similar to the entropy. The expectation value $\kl\td^2\kr$ can perfectly be computed for a given probability distribution $w(z,p)$. However, since $\td$ depends itself on $w$ we may call it a ``nonlinear observable''.
The statistical observable $P^2_s$ measures the ``roughness'' of the probability distribution in position space \cite{3A}.

A very interesting feature of the use of the classical wave function $\psi(x,p)$ and the quantum formalism is the possibility to express statistical observables as $P^2_s$ in the standard way as quantum operators. This allows us to use all the concepts from quantum mechanics, as the spectrum of the observables or the commutation relations and uncertainty relations. In the quantum formalism, $P_s$ is a perfectly acceptable observable. If it can be measured, the predictions for such measurements are the ones from quantum theory.

\section{Quantum particle as coarse grained classical particle}
\label{Quantumparticleas}

We have seen that both the classical and quantum particle admit a description in terms of a wave function within the quantum formalism. The wave functions depend on different arguments, however, $\psi_Q(x)$ for the quantum particle and $\psi_C(z,p)$ for the classical particle. In a sense, the classical particle uses more information since the wave function depends on  an additional momentum variable $p$. 

\medskip\noindent
{\bf 1. Quantum behavior from coarse graining}

This raises the question if the quantum particle can be obtained from the classical probability distribution $w(z,p)$ by a type of coarse graining, thereby discarding the additional information stored in the probability distribution for the classical particle. In the quantum formalism this coarse graining should involve a map from the classical density matrix $\rho_C(z,z',p,p')$ to the quantum density matrix $\rho_Q(x,x')$. In this section we construct an explicit realization of such a map. 
The coarse grained system can be considered as a subsystem of the classical statistical ensemble which describes the  particle in phase space. The part of the information contained in $\rho_Q$ characterizes the subsystem, while the information available from $w(z,p)$ which goes beyond $\rho_Q$ concerns the ``environment''.

We will see that only particular combinations of the classical observables (or operators) $\hat X_{cl},\ \hat P_{cl}$ and the statistical observables $\hat X_s,\hat P_s$ are compatible with the coarse graining, in the sense that the expectation values for position and momentum can be computed from $\rho_Q$. These are the quantum observables for position and momentum
\ba\label{17A}
\hat X_Q&=&\hat X_{cl}+\frac12\hat X_s,\nn\\
\hat P_Q&=&\hat P_{cl}+\frac12\hat P_s.
\ea
They obey standard quantum commutation relation
\be\label{113M}
[\hat X_Q,\hat P_Q]=i.
\ee

The discussion of the present section constitutes an explicit example how a quantum system can be obtained from a classical statistical ensemble, along the lines discussed in \cite{CWAA,CW2,CWE}. While in ref. \cite{3A} we express $\hat X_Q$ and $\hat P_Q$ as standard classical observables in an ensemble with a very large number of states, we express them here in terms of statistical observables in an ensemble characterized by the probability density in phase space. We find that these non-linear observables on the level of $w(z,p)$ are mapped onto linear observables on the level of the wave function $\psi_C(z,p)$.

The observation that a classical particle admits a description within the quantum formalism offers a convenient technical tool for the implementation of the coarse graining. For the quantum particle we will start with the same basic setting, namely the probability density $w(z,p)$ and associated wave function $\psi_C(z,p)$ in phase space. We can then use a standard tool for the description of subsystems in quantum physics, namely taking a subtrace of the density matrix. This averages over the states that are ``traced out'' or ``integrated out'', and therefore realizes a coarse graining in a simple way. The coarse grained system describes the quantum particle in the usual way and contains no longer any information beyond what is available in the usual quantum formalism.

As well known in quantum physics, a unitary time evolution of the original system does not necessarily induce a unitary time evolution of the subsystem. The subsystem will show a unitary time evolution only if its completely isolated from its environment - otherwise one may observe phenomena as decoherence \cite{DC} or syncoherence \cite{CW2}. We will encounter this feature in our coarse graining procedure if we use the Liouville equation for the time evolution of $w(z,p)$. Only for a free particle or for a quadratic potential the time evolution according to eqs. \eqref{H5}, \eqref{H8} is automatically unitary on the level of the coarse grained system.  As advocated in the introduction we will construct in the next sections a modification of the time evolution of the classical probability distribution $w(z,p)$ such that the coarse grained or quantum density matrix obeys a unitary time evolution for an arbitrary potential.

\medskip\noindent
{\bf 2. Coarse graining of density matrix}

Our starting point is the classical wave function $\psi_C(z,p)$. We perform first a Fourier transform, with $z=(x+y)/2$,
\be\label{QC2}
\tilde \psi_C(x,y)=\int_p e^{ip(x-y)}\psi_C
\left(\frac{x+y}{2},p\right).
\ee
In this ``position basis'' the classical wave function $\tilde\psi_C(x,y)$ depends on two commuting position variables $x$ and $y$. It is now a complex function which obeys
\be\label{33A}
\tilde\psi^*_C(x,y)=\tilde\psi_C(y,x).\ee
The Fourier transform \eqref{QC2} preserves the normalization which now reads
\be\label{20A}
\int_{x,y}|\psi(x,y)|^2=1.
\ee
The density matrix corresponding to this (pure state) classical wave function reads
\be\label{P3}
\tilde\rho_C(x,x',y,y')=\tilde\psi_C(x,y)\tilde\psi_C^*(x',y').
\ee

We next proceed to a ``coarse graining'' of the classical density matrix by ``tracing out'' the coordinates $(y,y')$, 
\be\label{P4}
\rho_Q(x,x')=\int_y\tilde\rho_C(x,x',y,y).
\ee
This coarse grained density matrix $\rho_Q(x,x')$ has all the properties of a quantum mechanical density matrix. One has
\be\label{P8}
\rho_Q^*(x',x)=\rho_Q(x,x')~,~\int_x\rho_Q(x,x)=1,
\ee
and the subtrace performed in eq. \eqref{P4} transfers the positivity of $\tilde\rho_C(x,x',y,y')$ to $\rho_Q(x,x')$. We will identify $\rho_Q(x,x')$ with the quantum density matrix of a quantum particle. 

A useful way to visualize the effect of the coarse graining is the decomposition of an arbitrary classical wave function $\tilde\psi_C(x,y)$ as 
\be\label{P9}
\tilde\psi_C(x,y)=\sum_{m,n}C_{nm}\chi_m(x)\zeta_n(y),
\ee
with some appropriate complete set of orthonormal basis vectors $\chi_m$ and $\zeta_n$ obeying
\be\label{P10}
\int_x\chi^*_{m'}(x)\chi_m(x)=\delta_{m'm}~,~\int_y\zeta^*_{n'}(y)\zeta_n(y)=\delta_{n'n}.
\ee
(For the special choice $\zeta_n=\chi^*_n$ the hermiticity condition \eqref{33A} implies $C^\dagger=C$.) In this basis the coarse grained density matrix reads
\be\label{P11}
\rho_Q(x,x')=\sum_{n,m,m'}C^*_{nm'}C_{nm}\chi^*_{m'}(x')\chi_m(x).
\ee

We can introduce positive normalized numbers $p_n$, 
\be\label{26A}
p_n=\sum_m|C_{nm}|^2\geq 0~,~\sum_n p_n=1,
\ee
where the normalization follows from eq. \eqref{20A}.
In terms of the normalized vectors $\varphi_n(x)$ defined by 
\ba\label{P12}
\sqrt{p_n}\varphi_n(x)&=&\sum_mC_{nm}\chi_m(x),\nn\\
\int_x\varphi^*_n(x)\varphi_n(x)&=&1
\ea
one finds
\be\label{P13}
\rho_Q(x,x')=\sum_n p_n\varphi^*_n(x')\varphi_n(x).
\ee
In general, $\rho_Q$ describes a mixed quantum state. Pure quantum states are special cases. As an example, they occur for $p_n=\delta_{n\bar n}$ such that only one term contributes in the sum \eqref{P13}. At this stage, the functions $\varphi_n(x)$ are not necessarily orthogonal to each other.
Interpreting $C_{nm}$ as elements of a matrix the normalization of $\tilde\psi_C(x,y)$ implies $\tr(C^\dagger C)=1$ and
\be\label{41A}
p_n=(CC^\dagger)_{nn},
\ee
with
\be\label{41B}
\sqrt{p_p p_n}\int_x\varphi^*_p(x)\varphi_n(x)=(CC^\dagger)_{np}.
\ee

The positivity of $\rho_Q$ can easily be established in this language. We want to show that for an arbitrary complex vector $f(x)$ one has
\ba\label{41C}
B&=&\int_{x,x'}f^*(x)\rho_Q(x,x')f(x')\geq 0,\nn\\
f(x)&=&\sum_m f_m\chi_m(x).
\ea
If $f$ is an eigenvector of $\rho$, $\int_{x'}\rho(x,x')f(x')=\lambda f(x)$,
this implies a positive eigenvalue $\lambda\geq 0$. Writing
\be\label{41D}
\rho_Q(x,x')=\sum_{m,m'} (C^\dagger C)_{m'm}\chi^*_{m'}(x')\chi_m(x)
\ee
one finds indeed
\be\label{32A}
B=\sum_{m,m'}f_{m'}(C^\dagger C)_{m'm}f^*_m=\sum_n b^*_n
b_n\geq 0,
\ee
with
\ba\label{41E}
b_n=\sum_m C_{nm}f^*_m.
\ea
This ends the proof that $\rho_Q(x,x')$ is a positive matrix, with all eigenvalues positive or zero. We mention that the choice of a coarse graining is, in principle, not unique. Possible alternatives are briefly discussed in appendix A.

\medskip\noindent
{\bf 3. Coarse grained observables}

The classical observables $A(z,p)$ are no longer available on the coarse grained level. 
Observables that can be computed in terms of $\rho_Q(x,x')$  can only involve $x$ and derivatives $\partial_x$. We can express such observables for the subsystem in terms of
the operators $\hat X_Q=x$ and $\hat P_Q=-i\partial_x$. An interesting question concerns the relation between these  observables on the coarse grained level and the original classical and statistical observables. 

Consider the expectation value
\ba\label{QC18}
\kl F_p(P_Q)\kr&=&\int_{x,x'}\delta(x-x')F_p(-i\partial_x)\rho_Q(x,x')\\
&=&\int_{x,x'} \delta(x-x')F_p\left(-\frac{i}{2}(\partial_x-\partial_{x'})\right)\rho_Q(x,x'),\nn
\ea
where the second equation uses the absence of boundary terms in the identity
\be\label{QC19}
\int_{x,x'}\delta(x-x')(\partial_x+\partial_{x'})
f(x,x')=\int_x\partial_xf(x,x)=0.
\ee
Using the definitions \eqref{P3}, \eqref{QC2} we evaluate this expectation value in terms of the original classical phase space probability distribution or associated phase space density matrix $\rho_C(z,z',p,p')$,
\be\label{43A}
\rho_C(z,z',p,p')=\psi_C(z,p)\psi_C(z',p'),
\ee
as
\ba\label{QC20}
&&\kl F_p(P_Q)\kr=\int_{x,x',y,y'}\delta(x-x')\delta(y-y')\nn\\
&&\hspace{1.9cm} F_p \left(-\frac i2(\partial_x-\partial_{x'})\right)
\rho(x,x',y,y')\nn\\
&&=\int_{p,p',x,x',y,y'}\delta(x-x')\delta(y-y')F_p
\left(-\frac i2(\partial_x-\partial_{x'})\right)\nn\\
&&\quad e^{ip(x-y)}e^{-ip'(x'-y')}
\rho_C\left(\frac{x+y}{2},\frac{x'+y'}{2},p,p'\right)\\
&&=\int_{p,z,z'}\delta(z-z')F_p
\left(p-\frac i4(\partial_z-\partial_{z'})\right)
\rho_C(z,z',p,p).\nn
\ea
We recognize the quantum operator $\hat P_Q$ in eq.~\eqref{17A},
\be\label{44A}
\hat P_Q=p-\frac{i}{2}\partial_z,
\ee
with 
\ba\label{QC21}
\kl F_p(P_Q)\kr&=&\int_{p,z,z'}
\delta(z-z')F_p\left(p-\frac i2\partial_z\right)
\rho_C(z,z'p,p)\nn\\
&=&\int_{p,z}\psi_C(z,p)F_p(\hat P_Q)\psi_C(z,p).
\ea
We will see below that this generalizes to arbitrary operators $F(x,-i\partial_x)$ acting on $\rho_Q(x,x')$. On the level of the phase-space density matrix $\rho_C(z,z',p,p')$ or the phase-space wave function $\psi_C(z,p)$ they are represented as  $F(\hat X_Q,\hat P_Q)$.

The expectation values of the quantum and classical momentum coincide, $\kl P_Q\kr =\kl p\kr$.  Nevertheless, we observe  an unusual fluctuation piece
\ba\label{QC24}
\kl P^2_Q\kr &=&\kl p^2\kr +\frac14\int_{z,p}\big (\partial_z\psi_C(z,p)\big)^2\nn\\
&=&\kl p^2\kr+\frac{1}{16}\int_{z,p}
\frac{\big(\partial_zw(z,p)\big)^2}{w(z,p)}\nn\\
&=&\Big \kl\Big(p+\frac14\partial_z\ln w(z,p)\Big)^2\Big\kr,
\ea
where the bracket on the r.h.s. denotes the classical expectation value 
$\kl A\kr=\int_{z,p}w(z,p)A(z,p)$. This adds a ``quantum dispersion'' to the momentum dispersion 
\ba\label{QC25}
\kl P^2_Q\kr-\kl P_Q\kr^2&=&\kl p^2\kr-\kl p\kr^2+\Delta^{(Q)2}_p,\nn\\
\Delta^{(Q)2}_p&=&\frac{1}{16}\int_{z,p}
\frac{\big(\partial_z w(z,p)\big)^2}{w(z,p)}.
\ea
Even if the classical momentum is sharp, 
$w(z,p)\sim\delta(p-\bar p)$, $\kl p^2\kr=\kl p\kr^2$, this holds for the quantum momentum only if $w(z,p)$ is independent of $z$. We see the uncertainty relation arising here in a direct way.

\medskip\noindent
{\bf 4. Quantum observables and statistical observables}

The quantum dispersion cannot be expressed by a linear classical observable of the usual type. It is rather related to the statistical observable
\be\label{49A}
\Delta_p^{(Q)2}=\frac{1}{4}\langle P^2_s\rangle.
\ee
It is interesting to see how a nonlinear observable for the classical particle appears as a linear observable on the coarse grained level. Indeed, the ``quantum momentum'' $\hat P_Q$  acts linearly on the classical wave function $\psi(z,p)$. Despite the appearance of the factor $i$ in the definition \eqref{44A} of $\hat P_Q$ the expectation value $\kl F_p(P_Q)\kr$ is always real. This follows from the hermiticity of $\hat P_Q$ if interpreted as a quantum operator in eq. \eqref{QC21}. Using partial integration one has for real $\psi(z,p)$ and integer $n$
\ba\label{188A}
&&\int_z\psi_C(z,p)\partial^{2n+1}_z\psi_C(z,p)=0,\\
&&\int_z\psi_C(z,p)\partial^{2n}_z\psi_C(z,p)=(-1)^n\int_z
\big (\partial^n_z\psi_C(z,p)\big)^2,\nn
\ea
such that the imaginary part of $\kl F_p(P_Q)\kr$ vanishes indeed. With $\partial_z\psi_C=(\psi_C/2)\partial_z\ln w$ we can express the second term in eq. \eqref{188A} by $wG(\partial^k_z\ln w)$, such that $\kl F(P_Q)\kr$ can indeed be expressed as a nonlinear expression in terms of $w$. 

Eq. \eqref{QC21} can be generalized to arbitrary expectation values of the type
\be\label{188B}
\kl F(X_Q,P_Q)\kr=\int_{p,z}\psi_C(z,p)F(\hat X_Q,\hat P_Q)\psi_C(z,p),
\ee
with
\be\label{52A}
\hat X_Q=z+\frac{i}{2}\partial_p.
\ee
This relates the expectation values of ``quantum observables'' for the coarse grained system to (nonlinear) classical observables which are computable in terms of the classical probability distribution $w(z,p)$.

We recognize the role of the statistical observables discussed in sect.~III. Indeed, we could have started with the operators $\hat X_Q$ and $\hat P_Q$ \eqref{113M} acting on the classical wave function $\psi_C(z,p)$
\ba\label{HA1}
\hat X_Q\psi_C(z,p)&=&\left (z+\frac{i}{2}\partial_p\right )\psi_C(z,p),\nn\\
\hat P_Q\psi_C(z,p)&=&\left (p-\frac i2\partial_z\right)\psi_C(z,p).
\ea
After a partial Fourier transform \eqref{QC2} these operators read
\ba\label{HA2}
\hat X_Q\tilde\psi_C(x,y)&=&x\tilde\psi_C(x,y)~,\nn\\
\hat P_Q\tilde\psi_C(x,y)&=&-i\partial_x\tilde\psi_C(x,y).
\ea
It is obvious that these particular operators can be implemented directly on the coarse grained level since they do not involve $y$. This provides an explanation why we have singled out the particular combinations \eqref{113M} as quantum observables.

We can now easily apply the quantum formalism on the coarse grained level, yielding
\ba\label{HA3}
\kl F(X_Q,P_Q)\kr&=&\int_{z,p}\psi_C(z,p)F(\hat X_Q,\hat P_Q)\psi_C(z,p)\nn\\
&=&\int_{x,x'}\delta(x-x')F(x,-i\partial_x)\rho_Q(x,x')\nn\\
&=&\int_x\psi^*_Q(x)F(\hat X_Q,\hat P_Q)\psi_Q(x),
\ea
where the last identity applies for a pure quantum state
and employs the operators $\hat X_Q,\hat P_Q$ in the usual position representation of quantum mechanics
\be\label{49B}
\hat X_Q=x\ ,\ \hat P_Q=-i\partial_x.
\ee

In particular, for the totally symmetrized correlation functions one obtains \cite{3A} in terms of the Wigner transform $\bar\rho_w(z,p)$ of $\rho_Q(x,x')$
\ba\label{HA4}
\kl F_s(X_Q,P_Q)\kr&=&\int_{z,p}\psi_C(z,p)F_s(\hat X_Q,\hat P_Q)\psi_C(z,p)\nn\\
&=&\int_{z,p}F_s(\hat X_Q,\hat P_Q)\bar \rho_w(z,p)\nn\\
&=&\int_{z,p}F_s(z,p)\bar\rho_w(z,p).
\ea
Here the Wigner transform \cite{Wig}, \cite{Moyal} of $\rho_Q$ is defined as $(z=(x+x')/2)$
\be\label{51}
\bar\rho_w(z,p)=\int d(x-x')e^{-ip(x-x')}\rho_Q(x,x')
\ee
and the operators $\hat X_Q,\hat P_Q$ defined in eq.~\eqref{17A} or eqs. \eqref{44A}, \eqref{52A} act linearly on $\bar\rho_w(z,p)$ in the second line of eq. \eqref{HA4}.

In summary, both the classical observables $X_{cl},P_{cl}$ and the quantum observables $X_Q,P_Q$ can be evaluated in terms of the classical probability distribution $w(z,p)$. While the classical observables obey the standard relation \eqref{2B} 
\be\label{HA5}
\kl F(X_{cl},P_{cl})\kr=\int_{z,p}F(z,p)w(z,p),
\ee
the quantum observables obey a similar relation \eqref{HA4} with $w$ replaced by $\bar\rho_w$. 

\medskip\noindent
{\bf 5. One particle distribution function}

The expectation values of symmetrized products of quantum operators can be found easily from the Wigner transform of the density matrix for the quantum particle according to eq. \eqref{HA4}.  We may therefore associate $\bar\rho_w(z,p)$ with the one particle distribution function for the quantum position and momentum.
It is interesting to compare the Wigner representation of the quantum density matrix $\bar\rho_w(z,p)$ with the classical probability distribution $w(z,p)$. With $z=(x+x')/2$ one finds from eqs. \eqref{51}, \eqref{P4}, \eqref{P3}, \eqref{QC2},
\ba\label{189A}
\bar\rho_w(z,p)=\int_{r,r',s,s'}\psi_C\left(z+\frac r2,p+s\right)\nn\\
\psi_C\left(z+\frac{r'}{2},p+s'\right)
e^{i(s'r-sr')}.
\ea
This yields eq. \eqref{AA1} in the introduction (with $z\to x$) since the imaginary part vanishes by the symmetry properties of the integrand. We find that $\bar\rho_w(z,p)$ and $w(z,p)$ differ, in general. In particular, it is possible that a positive $w(z,p)$ has a corresponding $\bar \rho_w(z,p)$ which becomes negative in certain regions of phase space.  Also the expectation values of symmetrized products of the quantum observables differ from the expectation values of the corresponding classical polynomials of $z$ and $p$ for the classical observables. We have already encountered a prominent example of this difference in eq. \eqref{QC24}. 

We may call eq. \eqref{189A} (or eq. \eqref{AA1}) the ``quantum transform'' of the classical probability distribution. The relative sign of $\psi_C(z,p)$ at two different points in phase space matters for $\bar\rho_w$, while it does not affect $w$. This raises the question if $\psi_C(z,p)$ contains more physical information than $w(z,p)$, or if the sign of $\psi_C$ is restricted by the properties of $w$ such that $\bar\rho_w$ can be expressed in terms of $w$ only. 
As we have discussed in ref.~\cite{3A}, the requirement that the continuity and differentiability properties of $w(z,p)$ are shared by $\psi_C(z,p)$ imposes severe constraints on the choice of sign of the classical wave function. This is closely related to the representation of statistical observables as derivative operators acting on $\psi_C$. 

To this discussion we add here a few more aspects. We have seen  that expectation values $\kl F(X_Q,P_Q)\kr$ can be expressed as appropriate functionals of $w(z,p)$ such that the sign of $\psi_C(z,p)$ never enters. Furthermore, if a sufficient number of correlations $F(X_Q,P_Q)$ are well defined such that $\bar\rho_w$ can be reconstructed from these correlations, we can conclude that $\bar\rho_w$ can indeed be computed from $w$ without the need of independent information about the sign of $\psi_C$. In other words, the restrictions on the sign of $\psi_C$ are strong enough in order to exclude any ambiguity of $\bar\rho_w$ for a given $w$.

If the time evolution of $\psi_C$ is determined by a first order differential equation (linear in $\partial_t$), the sign of $\psi_C(z,p,t)$ is fixed in terms of the sign of the initial wave function $\psi_C(z,p,0)$. This holds for arbitrary initial wave functions, including the case where $\psi_C(z,p,0)$ is negative in certain regions of phase space.  Thus for a given $\psi_C(z,p,0)$ the quantum  transform $\bar\rho_w(z,p,t)$ is, in principle, uniquely fixed for all $t$. Nevertheless, the possibility to express $\bar\rho_w(z,p,t)$ in terms of $w(z,p,t)$ (for the same $t$) tells us that no ``memory of the past'' is needed for the computation of the quantum transform of a classical probability distribution. In practice, it is most convenient to keep the sign information in $\psi_C(z,p,t)$ explicitly, however. In this sense it seems convenient to consider the real wave function $\psi_C(z,p)$ as the basic quantity which defines the classical ensemble in phase space. A few comments on the modification which would be introduced by a complex classical wave function can be found in appendix B. 

\medskip\noindent
{\bf 6. Quantum transform of classical wave packet}

It is instructive to consider as an explicit example the quantum transform of the wave packet 
\be\label{59A}
\psi(z,p)=(\Delta_x\Delta_p)^{-\frac{1}{2}}\exp\left\{-\frac{(z-\bar x)^2}{4\Delta^2_x}\right\}\exp\left\{-\frac{(p-\bar p)^2}{4\Delta^2_p}\right\},
\ee
where we use here a one-dimensional setting, with normalization
\be\label{59B}
\int\frac{dzdp}{2\pi}\psi^2(z,p)=1.
\ee
The integrals in eq. \eqref{189A} are Gaussian and we find
\ba\label{189B}
\bar\rho_w(z,p)&=&(\tilde\Delta_x\tilde\Delta_p)^{-1}
\exp \left\{-\frac{(z-\bar x)^2}{2\tilde\Delta^2_x}\right\}\nn\\
&&\exp-\left\{\frac{(p-\bar p)^2}{2\tilde\Delta^2_p}\right\},\nn\\
\tilde\Delta^2_x&=&\Delta^2_x+\frac{1}{16\Delta^2_p}~,~
\tilde\Delta^2_p=\Delta^2_p+\frac{1}{16\Delta^2_x},
\ea
to be compared with
\be\label{189C}
w(z,p)=(\Delta_x\Delta_p)^{-1}\exp 
\left\{-\frac{(z-\bar x)^2}{2\Delta^2_x}\right\}
\exp \left\{-\frac{(p-\bar p)^2}{2\Delta^2_p}\right\}.
\ee
We notice the difference between the dispersion of the quantum position $\tilde\Delta^2_x$ and the classical dispersion $\Delta^2_x$, 
\ba\label{189D}
\kl (X_Q-\bar x)^2\kr&=&\tilde \Delta^2_x=\Delta^2_x
\left(1+\frac{1}{16\Delta^2_x\Delta^2_p}\right),\nn\\
\kl(X_{cl}-\bar x)^2\kr&=&\Delta^2_x,
\ea
and similar for the quantum and classical momentum
\ba\label{189E}
&&\kl(P_Q-\bar p)^2\kr=\tilde \Delta^2_p=\Delta^2_p
\left(1+\frac{1}{16\Delta^2_x\Delta^2_p}\right),\nn\\
&&\kl(P_{cl}-\bar p)^2\kr=\Delta^2_p.
\ea
For the product of the quantum dispersions one obtains
\be\label{189F}
\tilde \Delta_x\tilde\Delta_p=\Delta_x\Delta_p+\frac{1}{16\Delta_x\Delta_p}.
\ee
For arbitrary $\Delta_x$ and $\Delta_p$ this expression has a minimum for $\Delta_x\Delta_p=1/4$. The value at the minimum obeys
\be\label{189G}
(\tilde\Delta_x\tilde\Delta_p)_{min}=\frac12.
\ee
This is precisely Heisenberg's uncertainty relation. We conclude that for a classical wavepacket, with classical dispersions $\Delta_x,\Delta_p$, the associated wave packet for the coarse grained quantum  particle has broader dispersions $\tilde\Delta_x,\tilde\Delta_p$. For a precisely located classical particle, $\Delta_x\to 0$, the quantum dispersion obeys $\tilde\Delta_x=1/(4\Delta_p)$, while $\tilde\Delta_p$ diverges.

We may also compute the Fourier transform \eqref{QC2} of the classical wave packet \eqref{59A},
\ba\label{189H}
\tilde\psi_C(x,y)=\left(\frac{\Delta_p}{\pi\Delta_x}\right)^{1/2}\exp 
\left\{-\frac{(x+y-2\bar x)^2}{16\Delta^2_x}\right\}\nn\\
\exp \{-\Delta^2_p(x-y)^2\}\exp \{i\bar p(x-y)\}.
\ea
For $\Delta_x\Delta_p=1/4$, corresponding to $\tilde\Delta_x\tilde\Delta_p=1/2$, one finds factorization
\ba\label{189I}
\psi(x,y)&=&\psi_Q(x)\psi^*_Q(y),\\
\psi_Q(x)&=&\left(\frac{4\Delta^2_p}{\pi}\right)^{1/4}
\exp \{-2\Delta^2_p(x-\bar x)^2\}
\exp \{i\bar px\}.\nn
\ea
The classical wave packets with $\Delta_x\Delta_p=1/4$ correspond to pure states of the associated quantum particle, with $2\Delta^2_p=1/(4\tilde\Delta^2_x)$. For $\Delta_x\Delta_p\neq 1/4$ one finds mixed quantum states. The mixed states teach us that the map from classical states to coarse grained quantum states is, in general, not injective. For given $\tilde \Delta_x,\tilde\Delta_p$ or given $\tilde r=\tilde\Delta_x/\tilde\Delta_p,\tilde s=\tilde \Delta_x\tilde\Delta_p$, we find 
$\Delta_x=(s\tilde r)^{1/2}$, $\Delta_p=(s/\tilde r)^{1/2}$, where $s=\Delta_x\Delta_p$ obeys eq. \eqref{189F}, $s+1/(16s)=\tilde s$. Except for the minimum at $s=1/4$ one finds two possibilities for $s$ for a given $\tilde s$. In particular, a large product of the dispersions, $\tilde s\gg1$,  can arise from either large $s\approx\tilde s$ or, in the opposite, from small $s\approx 1/(16\tilde s)\ll 1$. Very sharp classical states with $s\ll 1$ lead to broad quantum states with $\tilde s\gg1$ because of the dominant role of the contribution from statistical observables in this case. A narrow classical wave function has necessarily large gradients.

\medskip\noindent
{\bf 7. Quantum observables as probabilistic 

\hspace{0.2cm}observables}

A classical micro-state is specified by a point in phase space $(z,p)$. For each such state the classical observables $X_{cl}$ and $P_{cl}$ have a fixed value, namely $z$ and $p$. For the observables that can be defined on the coarse grained level this holds no longer true. In a given coarse grained state the quantum observables $X_Q$ and $P_Q$ have only a probability distribution of possible measurement values - they are ``probabilistic observables'' \cite{CWAA,CW2,CWE,3A,PO}. 

The states of the coarse grained system are denoted by points in the space of density matrices $\rho_Q(x,x')$. In a given state of the coarse grained system the observable $X_Q$ has the probability $w_Q(x)$ for finding the measurement value $x$. It is given by
\be\label{199A}
w_Q(x)=\rho_Q(x,x).
\ee
This can be seen by observing that the expectation values of arbitrary functions $F_x(X_Q)$ are determined by $\rho_Q(x,x)$, 
\be\label{199B}
\kl F_x(X_Q)\kr=\int_x F_x(x)w_Q(x)=\int_xF_x(x)\rho_Q(x,x).
\ee
Similarly, the probability $w_Q(p)$ to find for $P_Q$ the measurement value $p$ is given by
\be\label{199C}
w_Q(p)=\tilde\rho_Q(p,p),
\ee
with $\tilde \rho_Q(p,p')$ the Fourier transformation of $\rho_Q(x,x')$ with respect to both arguments. Again, one finds the relation
\be\label{66A}
\kl F_p(P_Q)\kr=\int_pF_p(p)w_Q(p).
\ee
We emphasize that $w_{Q}(x)$ and $w_{Q}(p)$ differ from the effective probabilities  
\be\label{A8}
w_C(x) = \int_p w (x,p),~ \  w_C(p) = \int_{x}w(x,p)
\ee
to find the classical position $x$ or classical momentum $p$ for a given classical statistical ensemble.

Eqs. \eqref{199B} and \eqref{66A} have important implications. They tell us that the rule \eqref{188B} for the computation of expectation values in terms of the classical wave function $\psi_C$ is compatible with interpretation of $X_Q$ and $P_Q$ as probabilistic observables for which the probability that a measurement of the quantum position finds a value $x$ is given by $w_Q(x)$. Indeed the discussion above implicitly assumes that there exist measurement devices that can measure $X_Q$, and that the expectation value for the product of $N$ measurements of $X_Q$ is given by $\kl X^N_Q\kr$ according to the rule \eqref{188B}. One can argue that this rule for the prediction of the outcome of possible measurements of $X_Q$ is not only consistent but also arises naturally. This requires, however, a more detailed description of the formal interpretation of a measurement process than presented here. 

The joint probability that $X_Q$ has the value $x$ and $P_Q$ has the value $p$ is not defined, neither on the level of the classical wave function $\psi(z,p)$ nor on the level of the coarse grained state $\rho_Q(x,x')$. In this sense we deal here with incomplete statistics \cite{3}, \cite{CWAA}. In contrast, we can define the conditional probability $w (p|x)$ of measuring $p$ for $P_Q$ if a measurement of $X_Q$ finds the value $x$. If we suppose that the measurement of $X_Q$ is compatible with the coarse graining we can compute the conditional probability $w(p|x)$ as 
\be\label{199D}
w(p|x)=\text{tr}\big\{\delta(P_Q-p)\rho_x\big\},
\ee
with $\rho_x$ a density matrix for which $X_Q$ has the sharp value $x$, namely
\be\label{199E}
\rho_x(x',x')=\delta(x'-x).
\ee
The condition \eqref{199E} is not yet sufficient for a determination of the density matrix $\rho_x(x'',x')$. 

Knowledge of the conditional probability $w(p|x)$ permits the computation of the ``conditional correlation'' or ``measurements correlation''
\be\label{199F}
\kl P_QX_Q\kr_m=\int_{p,x}pxw(p|x)w_Q(x).
\ee
With $w(p|x)w_Q(x)=\bar\rho_w(p,x)$ one finds that the conditional correlation is given by the quantum mechanical expectation value
\be\label{199G}
\kl P_QX_Q\kr_m=\frac12\kl\{P_Q,X_Q\}\kr.
\ee
For systems with a finite number of degrees of freedom an extensive discussion of conditional probabilities and their relation to quantum correlations can be found in ref.~\cite{CWAA}.

\medskip\noindent
\section {Time evolution of probability distribution}

Our basic setting for the description of a particle is the probability distribution $w(z,p)$. For a dynamical theory we have to formulate an ``evolution law'' for the time dependence of $w(z,p)$. This will replace Newton's laws for trajectories by a new fundamental dynamical law. The concept of trajectories and Newton's law will arise only as an effective description in an appropriate ``classical limit'' (which can be taken formally by $\hbar\to 0$). We will formulate the basic dynamical law as a linear first order differential equation for the wave
function $\psi_C(z,p)$. The form of this law is not known a priori. We will first investigate the evolution law for a classical particle which is given by the Liouville equation for  $w(z,p)$. Subsequently, we consider different forms of the basic evolution law, describing generalized particles.

\medskip\noindent
{\bf 1. Liouville equation for classical particle}

For the classical particle the time evolution is determined by the Liouville equation
\be\label{119A}
\partial_tw(z,p)=\left\{-\frac pm\partial_z+\frac{\partial V(z)}{\partial z}\partial_p\right\}
w(z,p).
\ee
The Liouville equation for the classical wave function $\psi_C(z,p)$ can formally be written as a 
Schr\"odinger equation for a pure state of a classical particle, which is equivalent to the Liouville equation
\be\label{QC1}
i\partial_t\psi_C(z,p)=H_L\psi_C(z,p)=
\left(-i\frac pm\partial_z+iV'(z)\partial_p\right)\psi_C(z,p).
\ee
Using this dynamic equation all probabilistic features of one classical particle in a potential can be described. This extends to non-interacting multi-particle systems. In the position basis this generalized Schr\"odinger equation reads 
\ba\label{QC3}
i\partial_t\tilde\psi_C(x,y)&=&H_L\tilde\psi_C(x,y),\\
H_L&=&-\frac{1}{2m}(\partial^2_x-\partial^2_y)+V'
\left(\frac{x+y}{2}\right)(x-y).\nn
\ea

\medskip\noindent
{\bf 2. Time evolution of coarse grained density matrix}

According to eq. \eqref{QC3} the time evolution of the coarse grained density matrix obeys
\ba\label{P5}
i\partial_t\rho_Q(x,x')&=&\left\{-\frac{1}{2m}(\partial^2_x-\partial^2_{x'}+
V(x)-V(x')\right\}\rho_Q(x,x')\nn\\
&&-E(x,x'),
\ea
with
\be\label{P6}
E(x,x')=\int_y\big\{\tilde\psi_C^*(x',y){\cal H}_x\tilde\psi_C(x,y)-\tilde\psi_C(x,y)
{\cal H}_{x'}\tilde\psi_C^*(x',y)\big\}.
\ee
Here we have decomposed $H_L$ into a part $H_Q$ acting only on $x$ and an $x$-dependent part ${\cal H}_x$ acting on $y$,
\ba\label{P2}
H_L&=&H_Q-{\cal H}_x,\nn\\
H_Q&=&-\frac{1}{2m}\partial^2_x+V(x),\nn\\
{\cal H}_x&=&-\frac{1}{2m}\partial^2_y+W(x,y),
\ea
with
\be\label{P1}
W(x,y)=V'\left(\frac{x+y}{2}\right)(y-x)+V(x).
\ee
The sum of the contributions from the derivative terms $\sim \partial^2_y$ in ${\cal H}_x$ vanishes by partial integration, such that
\ba\label{P7}
E(x,x')&=&\int_y\psi^*(x',y)\big(W(x,y)-W(x',y)\big)\psi(x,y)\nn\\
&=&-E^*(x',x),\nn\\
E(x,x)&=&0.
\ea
For $E(x,x')=0$ eq. \eqref{P5} would describe the von-Neumann equation for the evolution of the density matrix for a quantum particle in a potential $V$. In general, $E(x,x')$ does not vanish, however.

\medskip\noindent
{\bf 
3. Conditions for unitary time evolution from 

\hspace{0.2cm}Liouville equation}

For an unharmonic potential $V$ and the most general classical $\psi_C(z,p)$ eq. \eqref{P5} does not describe a unitary evolution. This is of no surprise, since a reduction of a density matrix to a subsystem by tracing out degrees of freedom is generally used in quantum  physics for a description of decoherence or for the increase of entropy for the subsystem. It is nevertheless interesting to ask under what conditions the time evolution of the quantum particle described by $\rho_Q(x,x')$ remains unitary. We will associate a unitary evolution with an effective ``isolation'' of the subsystem in the sense that its time evolution is deconnected from its ``environment'' and can be described by properties of the subsystem alone. Here the properties of the subsystem are uniquely characterized by the quantum density matrix $\rho_Q$, whereas all information in $\tilde\psi_C(x,y)$ beyond $\rho_Q$ is ascribed to properties of the environment. The general ``isolation condition'' or ``unitarity condition'' for a unitary time evolution of $\rho_Q$ requires the evolution equation to be of a Hamiltonian form, or
\ba\label{T1}
E(x,x')&=&-\int_w\big\{S(x,w)\rho_Q(w,x')-\rho_Q(x,w)S(w,x')\big\},\nn\\
S(x,w)&=&S^*(w,x).
\ea
In this case the evolution of $\rho_Q$ obeys the unitary evolution according to the van Neumann equation
\be\label{T2}
i\partial_t\rho_Q=\big[\bar H_Q,\rho_Q],
\ee
with quantum Hamiltonian
\be\label{T3}
\bar H_Q=\delta(x-w)
\left(-\frac{1}{2m^2}\partial^2_w+V(w)\right)+S(x,w).
\ee

Of particular interest is a local evolution with 
\be\label{T4}
S(x,w)=-\delta(x-w)\epsilon(x).
\ee
In this case the classical probability distribution has to obey the ``locality condition''
\be\label{T5}
E(x,x')=\big[\epsilon(x)-\epsilon(x')\big]\rho_Q(x,x').
\ee
This is stronger than the unitarity condition. Using eq. \eqref{P7} we can write the locality condition as an integral condition for the classical density matrix
\ba\label{T6}
\int_y\big[W(x,y)-W(x',y)\big]\tilde\rho_C(x,x',y,y)\nn\\
=\big[\epsilon(x)-\epsilon(x')\big]\int_y\tilde\rho_C(x,x',y,y).
\ea
A classical probability distribution which obeys for all times the locality condition \eqref{T6} describes a quantum particle moving in a ``quantum potential''
\be\label{T7}
V_Q(x)=V(x)-\epsilon(x).
\ee
The coarse graining results in this case in a modification of the effective potential seen by the particle by an $x$-dependent ``quantum shift'' $\epsilon(x)$. 

Before discussing the locality condition \eqref{T6} in more detail we investigate first a classical particle moving in a quadratic classical potential 
\be\label{T8}
V(x)=a+bx+\frac c2 x^2.
\ee
This will give a simple existence proof that a unitary time evolution is possible for our construction of a quantum particle as a coarse grained classical particle. For a quadratic  classical potential the associated $W(x,y)$ in eq. \eqref{P1} does not depend on $x$
\be\label{T9}
W(x,y)=a+by+\frac c2 y^2.
\ee
The locality condition \eqref{T6} is obeyed for all classical density matrices with $\epsilon(x)=0$. The quantum potential $V_Q(x)$ coincides with the classical potential $V(x)$ in this case. In particular, a product form of the classical wave function
\be\label{T10}
\tilde\psi_C(x,y)=\chi(x)\zeta(y)~,~\int_x|\chi(x)|^2=\int_y|\zeta(y)|^2=1,
\ee
leads to a pure quantum state
\be\label{T11}
\rho_Q(x,x')=\chi(x)\chi^*(x').
\ee

\medskip\noindent
{\bf 4. Locality condition for the Liouville equation}

The issue of the time evolution gets more involved for unharmonic classical potentials. As an example we may consider a quartic classical potential
\ba\label{N1}
V&=&\frac c2 x^2+\frac\lambda8 x^4,\\
W&=&\frac c2 y^2+\frac{\lambda}{16}
(y^4+2xy^3-2x^3y+x^4),\nn
\ea
for which the locality condition \eqref{T6} reads

\ba\label{N2}
&&\frac{\lambda}{16}
\int_y\big\{2(x-x')y^3-2(x^3-x'^3)y+(x^4-x'^4)\big\}\nn\\
&&\tilde\rho_C(x,x',y,y)\nn\\
=&&\big[\epsilon(x)-\epsilon(x')\big]
\int_y\tilde\rho_C(x,x',y,y).
\ea

A solution of the locality condition \eqref{T6} by itself is not difficult. For example, we may take for $\psi(x,y)$ a solution of the stationary Schr\"odinger equation
\be\label{N3}
\big(\hat H_y+W(x,y)\big)\tilde\psi_C(x,y)=\epsilon(x)\tilde\psi_C(x,y),
\ee
where $x$ is kept fixed, $y$ is considered as the variable and $\hat H_y$ is a hermitean operator obeying
\ba\label{N4}
\int_y\tilde\psi_C^*(x',y)\hat H_y\tilde\psi_C(x,y)=\int_y\hat H^*_y\tilde\psi_C^*(x',y)\tilde\psi_C(x,y)
\ea
(For $\hat H_y=-\partial^2_y/2m$ eq. \eqref{N3} reads 
${\cal H}_x\tilde\psi_C=\epsilon(x)\tilde\psi_C$.) Since the effective potential in eq. \eqref{N3} depends on $x$ through $W(x,y)$, also the eigenvalues of $\hat H_y+W(x,y)$, depend on $x$. For every possible eigenvalue $\epsilon(x)$ we have a corresponding solution of eq. \eqref{N3}, and therefore a wave function which obeys the locality constraint. Furthermore, there are many different choices of $\hat H_y$ compatible with eq. \eqref{N4}. (For example, $\hat H_y$ can contain an arbitrary potential $\Delta V(y)$.) Typically, $\epsilon(x)$ depends on the choice of $\hat H_y$. We also note that for a given $x$ the solution of eq. \eqref{N3} is not normalized, with
\be\label{N5}
\int_y|\tilde\psi_C(x,y)|^2=\rho_Q(x,x).
\ee
Since $\hat H_y$ does not contain $x$-derivatives the wave function $\hat\psi(x,y)=\chi(x)\tilde\psi_Q(x,y)$ is also a solution of eq. \eqref{N3}. As another example, any $\tilde\rho_C(x,x',y,y)$ which is symmetric in $y\to -y$ solves the particular locality condition \eqref{N2}, with $\epsilon=\lambda x^4/16$. 

The crucial issue concerns the question if the locality condition \eqref{T6} holds for all times. A priori, it is not guaranteed that an initial wave function which obeys eq. \eqref{N3} at some $t_0$ has a time evolution according to eq. \eqref{QC3} such that the locality condition continues to hold for $t>t_0$. If we concentrate on solutions of eq. \eqref{N3}, this would require that $\partial_t\tilde\psi_C(x,y)$ and higher time derivatives of $\tilde\psi$ also obey the differential equation \eqref{N3}, with the same choice of $\hat H_y$ and $\epsilon(x)$.

Let us assume that at some initial time $t_0=0$ the probability distribution obeys the locality condition with a suitable $\epsilon(x)$. For $t>0$ the coarse grained density matrix obeys
\ba\label{205A}
&&\rho_Q(x,x';t)=\\
&&\int_{y,y'}\delta(y-y')e^{-iH_L(x,y)t}
e^{iH_L(x',y')t}
\tilde\rho_C(x,x',y,y';0).\nn
\ea
On the other hand, if the locality condition holds for all $t$ one can also infer
\be\label{205B}
\rho_Q(x,x';t)=
e^{-i\bar H_Q(x)t}
e^{i\bar H_Q(x')t}
\rho_Q(x,x';0).
\ee
Since $\bar H_Q$ and $H_L$ do not commute for unharmonic potentials, it is not easy to reconcile eq. \eqref{205A} with eq. \eqref{205B}. In general, a factorized form $\tilde\psi_C(x,y)=\chi(x)\zeta(y)$ is not preserved by eq. \eqref{205A}, nor are particular symmetry properties as $\tilde\psi_C(x,y)=\pm\tilde\psi_C(x,-y)$. At the present stage we do not know if initial distributions $\tilde\rho_C(x,x',y,y';0)$ exist such that eq. \eqref{205B} holds for all $x,x',t$ with a suitable $\epsilon(x)$. Initial distributions of this type would lead to a unitary time evolution on the coarse grained level even if the Liouville equation is used as the basic dynamic equation. 

\medskip\noindent
{\bf 5. Generalized  particles}

The issue of a unitary evolution for a coarse grained subsystem depends crucially on the Hamiltonian which specifies the time evolution of the probability distribution for the classical particle. We may consider a generalized notion of particles for which the time evolution of the probability distribution $w(z,p)$ may be different from the Liouville equation. Since we do not assume an underlying deterministic behavior of point particles (as for the Liouville equation) there is, a priori, substantial freedom in the choice of an evolution law for $w(z,p)$. We will assume that it follows from a first order differential equation. We will see that for a suitable modification of the time evolution of $w(z,p)$ it is straightforward to obtain a unitary time evolution of the coarse grained system. 

We will replace the Hamiltonian $H_L$ in eq. \eqref{QC3} by 
\be\label{GP1}
H_W=-\frac{1}{2m}(\partial^2_x-\partial^2_y)+V(x)-V(y),
\ee
such that the classical wave function obeys
\be\label{GP1QA}
\partial_t\tilde\psi_C(x,y)=-iH_W\psi(x,y).
\ee
Since $H_W$ is hermitean the time evolution of the wave function is unitary and the normalization of the classical probabilities is preserved
\be\label{GP1B}
\partial_t\int_{z,p}w(z,p)=\partial_t\int_{x,y}
|\tilde\psi_C(x,y)|^2=0.
\ee
Also $w(z,p)=\psi^2_C(z,p)$ is manifestly positive semidefinite for all $t$. Eqs. \eqref{GP1QA}, \eqref{GP1} constitute the new basic dynamical equation for the evolution of the probability density in phase space. We will see that this evolution equation describes a quantum particle in a potential $V$. For a free particle or a quadratic potential $H_W$ coincides with $H_L$.

We can write $H_W=H_Q-\tilde H_Q$, where $H_Q$ and $\tilde H_Q$ depend only on $x$ and $y$, respectively, with $[H_Q,\tilde H_Q]=0$. Adapting eq. \eqref{P2} to the modified Hamiltonian implies $W(x,y)=V(y)$, such that $E(x,x')=0$ in eq. \eqref{P7} and the locality condition \eqref{T5} is obeyed automatically, with $\epsilon(x)=0$. For a generalized particle with time evolution given by $H_W$ the coarse grained quantum particle always follows a unitary evolution with Hamiltonian $H_Q$. The discussion of the relation between the quantum operators $\hat X_Q$ and $\hat P_Q$ and the commuting classical observables $\hat X_{cl},\hat P_{cl}~(=z,p)$ does not depend on the choice of the Hamiltonian and therefore remains valid for the generalized particle as well. This also holds for the one particle distribution function. 

It is interesting to express the Hamiltonian $H_W$ in terms of the standard position and momentum variables $z,p$ in phase space:
\be\label{GP2}
H_W=-i\frac pm\partial_z+V\left(z+\frac i2\partial_p\right)-V
\left(z-\frac i2\partial_p\right).
\ee
For unharmonic potentials this involves higher momentum derivatives $\partial^n_p$. For example, for the quartic potential \eqref{N1} one finds
\be\label{GP3}
H_W=H_L-i\frac\lambda8 z\partial^3_p.
\ee
We observe that the generalization of the Liouville operator $L_W=iH_W$ remains a real operator, but no longer linear in the derivatives. A real classical wave function $\psi_C(z,p)$ remains real for all $t$. This implies an evolution equation for $w(z,p)$
\ba\label{GP3A}
\partial_t w(z,p)&=&2\psi_C(z,p,t)\partial_t\psi_C(z,p,t)=-\hat L w+C[w],\nn\\
C[w]&=&-2\psi_C \left (iV\left (z+\frac i2\partial_p\right)
-iV\left(z-\frac i2\partial_p\right)\right.\nn\\
&&\left.+\frac{\partial V}{\partial z}(z)\partial_p\right)\psi_C,
\ea
where we can use $\partial_p\psi_C=(\psi_C/2)\partial_p\ln w$. 

In summary, the time evolution of the probability density in phase space obeys for a generalized particle
\be\label{M2A}
\partial_t w=-2\sqrt w L_W \sqrt w, 
\ee
with
\be\label{M2B}
L_W=\frac p m \partial_z + \frac i \hbar V \left( z + \frac{i\hbar} 2 \partial_p    \right) - \frac i \hbar V \left( z - \frac{i\hbar} 2 \partial_p  \right).
\ee
Writing $\hbar$ explicitly in eq. \eqref{M2B} demonstrates that in the classical limit for $\hbar \to 0$ one recovers $L_W\to\hat L$. Thus $C[w]$ in eq. \eqref{GP3A} can be interpreted as the quantum modification of the dynamical law for the time evolution of phase-space probabilities. For the quartic potential \eqref{N1}, \eqref{GP3} this yields the non-linear modification of the Liouville equation for the probability density
\be\label{GP3B}
C[w]=-\frac{\lambda \hbar^2} {8}z
\left(\partial^3_p\ln w+\frac 32\partial^2_p\ln w\partial_p\ln w+
\frac14(\partial_p\ln w)^3\right)w.
\ee
We can generalize this to an arbitrary form of a potential $V(x)$ which can be written in terms of an expansion in powers of $x$. Eqs. \eqref{M2A}, \eqref{M2B} may be considered as the new basic dynamic equation for the time evolution of the probability density in phase space. Despite the non-linearity of eq. \eqref{M2A} on the level of $w$ we recall that the evolution equation for the phase-space wave function $\psi_C$ remains a linear differential equation. 

\medskip\noindent
\section{Quantum particle from classical probabilities}
\label{Quantum particle from classical probabilities}

If the time evolution of $w(z,p)$ follows eq. \eqref{GP3A} all features of a quantum particle in a potential will be found for the coarse grained subsystem. This is a direct demonstration that a suitable time evolution of a classical ensemble, with classical probability distribution $w(z,p)$, can describe a quantum system as an appropriate subsystem \cite{CWAA,CW2,CWE,3A}. The relation between the non-commuting quantum operators on one side and the classical and statistical observables on the other side has been discussed explicitly above.

\medskip\noindent
{\bf 1. \quad Quantum pure states}

A crucial concept for the understanding of quantum physics concerns the quantum pure states $\psi_Q(x)$. (We use here the position basis.) To each quantum pure state we can associate a classical wave function which takes a simple product from in the position basis
\be\label{103A}
\tilde \psi_C(x,y)=\psi_Q(x)\psi^*_Q(y).
\ee
The Fourier transform of eq. \eqref{103A} yields for pure quantum states the simple identification of the classical wave function and the Wigner function
\be\label{103B}
\psi_C(z,p)=\bar\rho_w(z,p),
\ee
and, in consequence, the expression of the probability in phase space as the square of the Wigner function
\be\label{102C}
w(z,p)=\bar\rho^2_w(z,p).
\ee
One may check that for the special case of pure quantum states the identification \eqref{103B} is compatible with the general definition of $\bar\rho_w$ by eq. \eqref{189A}. 

It is straightforward to verify that a classical wave function with the product form \eqref{103A} yields, after coarse graining, a pure state density matrix. Insertion of eq. \eqref{103A} into the definition \eqref{P4}, \eqref{P3} yields indeed 
\be\label{103D}
\rho_Q(x,x')=\psi_Q(x)\psi^*_Q(x'),
\ee
whereby the normalization condition $\int_y|\psi_Q(y)|^2=1$ assures that $\rho_Q$ is indeed quadratic (and not quartic) in $\psi_Q$. The proof that the choice \eqref{103A} for $\tilde \psi_C(x,y)$ is unique for every pure quantum state $\psi_Q(x)$ proceeds along the following line: Quantum pure states can only be described by a classical wave function which factorizes in the $(x,y)$ basis
\ba\label{GP4A}
\tilde\psi_C(x,y)=\psi_Q(x)\zeta(y)~,~\int_y\zeta^*(y)\zeta(y)=1.
\ea
For a real classical wave function $\psi_C(z,p)$ the corresponding transform $\tilde\psi_C(x,y)$ must be hermitean, $\tilde\psi_C^*(x,y)=\tilde\psi_C(y,x)$, implying $\zeta(y)=\psi^*_Q(y)$. 

In order to show the factorization property \eqref{GP4A} we employ the representation of a pure state density matrix
\ba\label{104A}
\psi_Q(x)&=&\sum_mb_m\chi_m(x),\nn\\
\rho_Q(x,x')&=&\sum_{m,m'}b_mb^*_{m'}\chi_m(x)\chi^*_{m'}(x').
\ea
Comparison with the expansion \eqref{P9} of $\tilde\psi_C(x,y)$ and definition of the coarse grained density matrix \eqref{P4}, \eqref{P11} yields the condition
\be\label{104B}
(C^\dagger C)_{m'm}=b^*_{m'}b_m.
\ee
By a suitable unitary transformation
\be\label{104C}
C=U^\dagger\tilde C U~,~b=U^T\tilde b,
\ee
one obtains $\tilde b_m=\delta_{mk}$ and 
\be\label{104D}
(\tilde C^\dagger \tilde C)_{m'm}=\delta_{m'k}\delta_{mk},
\ee
with arbitrary fixed $k$, e.g. $k=1$. The general solution of eq. \eqref{104D} is 
\be\label{104E}
\tilde C_{nm}=\tilde U_{nk}\delta_{mk}
\ee
with $\tilde U$ an arbitrary unitary matrix. In this basis $\tilde C_{nm}$ factorizes
\be\label{104F}
\tilde C_{nm}=\tilde a_n\tilde e_m~,~\tilde a_n=\tilde U_{nk}~,~\tilde e_m=\delta_{mk}.
\ee
The factorization property is preserved by unitary transformations. Therefore $C$ factorizes 
\ba\label{104Ga}
C_{nm}=a_ne_m
\ea
and
\ba\label{104G}
\tilde \psi_C(x,y)&=&\chi(x)\zeta(y),\\
\chi(x)&=&\sum_na_n\chi_n(x)~,~\zeta(y)=\sum_me_m\zeta_m(y).\nn
\ea
Using the fact that $\zeta(y)$ obeys the normalization \eqref{GP4A} we finally obtain $\chi(x)=\psi_Q(x)$ by insertion into eq. \eqref{P4}. This closes the proof the factorized classical wave function \eqref{103A} is unique for a pure quantum state.

It is remarkable that for pure quantum states the classical wave function $\psi_C$ is fixed uniquely. In general, some information is lost in a coarse graining procedure as in eq. \eqref{P4}. This remains true in our case, since the solution of the coarse graining condition still allows for an arbitrary normalized function $\zeta(y)$. The additional condition of hermiticity \eqref{33A} is sufficiently strong, however, such that $\zeta(y)$ is fixed uniquely. (More precisely, the hermiticity condition
\be\label{104H}
\psi^*_Q(x)\zeta^*(y)=\zeta(x)\psi_Q(y)
\ee
still allows for an overall sign $\zeta(y)=\pm \psi^*_Q(y)$. This only concerns the overall sign of $\tilde \psi_C$ and $\psi_C$ and leaves $w=\psi^2_C$ unique. The sign ambiguity for neighboring time arguments is fixed by the analyticity requirement of the time evolution.)

\medskip\noindent
{\bf 2. Static classic wave functions for pure quantum states}

It is interesting to study the special case of static classical wave functions $\psi_C(z,p)$ which correspond to pure quantum states
\ba\label{GP4B}
i\partial_t\tilde\psi_C(x,y)&=&(H_Q-\tilde H_Q)\tilde\psi_C(x,y)\\
&=&\big (H_Q\psi_Q(x)\big)\psi^*_Q(y)-\psi_Q(x)\big(\tilde H_Q\psi_Q(y)\big)^*\nn\\
&=&\big (i\partial_t\psi_Q(x)\big)\psi^*_Q(y)-\psi_Q(x)\big(i\partial_t\psi_Q(y)\big)^*=0.\nn
\ea
Eq. \eqref{GP4B} is obeyed for any eigenvector of $H_Q$
\be\label{GP4C}
H_Q\psi^{(n)}_Q(x)=E_n\psi^{(n)}_Q(x),
\ee
since eq. \eqref{GP4C} implies
\be\label{GP4D}
\tilde H_Q\psi^{(n)}_Q(y)=E_n\psi^{(n)}_Q(y).
\ee
We find that stationary states of the quantum Schr\"odinger equation correspond to time independent classical probability distributions. More generally, a static classical probability distribution $w(z,p)$ is mapped to a static Wigner representation of the density matrix $\bar\rho_w(z,p)$ by the quantum transform \eqref{AA1}.

\medskip\noindent
{\bf 3.\quad Quantum energy}

This discussion shows that the Hamiltonian $H_W$ in eq. \eqref{GP1} cannot be associated to the energy observable for the quantum particle. Instead, $H_W$ commutes with the energy observable
\be\label{GP4E}
\tilde E=\frac12(H_Q+\tilde H_Q),
\ee
which is therefore a conserved quantity. After coarse graining, this energy observable has the properties of the energy in quantum mechanics. For the stationary pure state obeying eqs. \eqref{GP4C}, \eqref{GP4D} one finds
\be\label{GP4F}
\tilde E\tilde\psi_C(x,y)=E_n\tilde\psi_C(x,y).
\ee\\
More generally, the expectation value of $\tilde E$ can be expressed in terms of the coarse grained density matrix \eqref{P4}
\ba\label{GP4G}
\kl \tilde E\kr&=&\frac12\int_{x,y}\tilde\psi_C^*(x,y)(H_Q+
\tilde H_Q)\tilde\psi_C(x,y)\nn\\
&=&\int_{x,y}\tilde\psi_C^*(x,y)H_Q\tilde\psi_C(x,y)\\
&=&\int_{x,x'}\delta(x-x')H_Q\rho_Q(x,x')=\text{tr}(H_Q\rho_Q),\nn
\ea
where the second line uses the hermiticity of $\tilde\psi_C(x,y)$. Eq. \eqref{GP4G} is the standard expression for the quantum mechanical energy in terms of the density matrix.

\medskip\noindent
{\bf 4. \quad Quantum particle}

In summary, a classical probability distribution $w(z,p)=\psi_C^2(z,p)~,~\psi_C=\psi_C^*$, leads to a coarse grained one particle distribution function
\ba\label{GP5}
\bar\rho_w(z,p)=&&\int_{r,r's,s'}\psi_C
\left(z+\frac r2,p+s\right)
\psi_C\left(z+\frac{r'}{2},p+s'\right)\nn\\
&&\cos(s'r-sr').
\ea
This allows the computation of expectation values and totally symmetrized correlation functions of coarse grained position and momentum observables $X_Q,P_Q$ according to 
\be\label{GP6}
\kl F_s(X_Q,P_Q)\kr=\int_{z,p}F(z,p)\bar\rho_w(z,p).
\ee
They are identical to the expectation values of totally symmetrized products of non-commuting quantum operators $\hat X_Q,\hat P_Q$. The coarse grained observables $X_Q,P_Q$ can be related to the classical observables $X_{cl},P_{cl}$ whose expectation values are computed in terms of $w$
\be\label{GP7}
\kl F(X_{cl},P_{cl})\kr=\int_{z,p}F(z,p)w(z,p).
\ee
This relation involves, however, non-linear expressions of $w(z,p)$ for the expectation values of statistical observables as
\be\label{GP8}
\kl P^2_Q\kr=\kl P^2_{cl}\kr+\frac{1}{16}\kl (\partial_z\ln w)^2\kr.
\ee

If  the time evolution of the classical probability distribution is determined by 
\ba\label{GP9}
\partial_t\psi_C&=&-\frac pm\partial_z\psi_C\\
&&-i\left[V(\left(z+\frac i2\partial_p\right)-V
\left(z-\frac i2\partial_p\right)\right]\psi_C,\nn
\ea
the coarse grained one particle distribution evolves according to the unitary evolution of a quantum particle in a potential $V(x)$. The evolution of $\psi_C$ is given by a simple linear equation, while non-linearities appear in the corresponding evolution of $w$. 

A pure state of the quantum particle arises from a classical probability distribution for which $\psi_C(z,p)$ obeys the factorization property
\ba\label{GP10}
\int_p e^{ip(x-y)}\psi_C
\left(\frac{x+y}{2},p\right)=
\psi_Q(x)\psi^*_Q(y).
\ea
The quantum wave function $\psi_Q(x)$ is complex, in general. For $\psi_C(z,p)$ evolving according to eq. \eqref{GP9} the factorization condition \eqref{GP10} is preserved for all $z$ and the time evolution of $\psi_Q(x)$ follows the usual Schr\"odinger equation. The expectation values for quantum operators $\hat X_Q,\hat P_Q$ obey the usual quantum rules. 

\medskip\noindent
{\bf 5. \quad Quantum time evolution}

Using the definition \eqref{GP5} we can derive from eq. \eqref{GP9} the time evolution of the Wigner function
\ba\label{119B}
&&\bar\partial_t\rho_w(z,p)=
\left\{-\frac pm\partial_z-iV\left(z+\frac i2\partial_p\right)\right.\nn\\
&&\qquad~~ \qquad \left.+iV\left(z-\frac i2\partial_p\right)\right\}\bar\rho_w(z,p)\\
&&=\left\{-\frac pm\partial_z+2 V(z)\sin\left(\frac 12
\stackrel{\leftarrow}{\partial_z} \stackrel{\rightarrow}{\partial_p}\right)\right\}
\bar\rho_w(z,p).\nn
\ea
This coincides, perhaps somewhat surprisingly at first sight in view of the folding in eq. \eqref{GP5}, with the form of the evolution of the classical wave function \eqref{AA2}. It is well known that eq. \eqref{119B} is equivalent to the time evolution of the density matrix for the quantum particle
\be\label{119C}
\partial_t\rho_Q=-i[H_Q,\rho_Q],
\ee
with
\be\label{119D}
H_Q=\frac{\hat P^2_Q}{2m}+V(\hat X_Q).
\ee
Eq. \eqref{119C} yields, for $z=(x+x')/2$, the relation
\ba\label{119E}
&&\partial_t\rho_Q(x,x')=\int_p\bar\partial_t\rho_w(z,p)e^{ip(x-x')}\\
&&=\int_p\left\{-iV(x)+iV(x')-
\frac pm\partial_z\right\}
\bar\rho_w(z,p)e^{ip(x-x')},\nn
\ea
from which one infers eq. \eqref{119B} by partial integration. For a formal Taylor expansion of $V$ in eq. \eqref{119B} in $i\partial_p$ one finds
\be\label{119F}
\partial_t\bar\rho_w(z,p)=
\left\{-\frac pm\partial_z+\frac{\partial V}{\partial z}\partial_p-\frac{1}{24} 
\frac{\partial^3 V}{\partial z^3}\partial^3_p+\dots\right\}
\bar\rho_w(z,p)
\ee

\section{Classical position and momentum for quantum particle}
\label{Classicalposition}

In the preceding section we have constructed a map from a classical probability distribution in phase space $w(z,p)$ to the Wigner transform of the density matrix for a quantum  particle $\bar\rho_w(z,p)$. Inversely, one may ask if one can find a suitable $w(z,p)$ for every quantum state. In this section we show that this is indeed the case. For any given pure or mixed quantum state we construct a classical probability distribution whose quantum transform corresponds to the Wigner transform of the density matrix for this state. We find that $w(z,p)$ is unique for pure quantum states. For a given $w(z,p)$ the expectation values of the classical observables $F(X_{cl},P_{cl})$ can be evaluated. This allows us to define classical observable also for pure quantum states. For mixed quantum states the definition of classical observables may require additional information, since $w(z,p)$ may not be unique for a given $\bar\rho_w(z,p)$ in this case. 

\medskip\noindent
{\bf 1. Pure quantum states}

Consider first the wave function $\psi_Q(x)$ for a pure state of a quantum  particle. According to eq. \eqref{103B} the classical wave function equals the quantum density matrix in this case, and therefore $\tilde \psi_C$ equals the quantum mechanical density matrix $\rho_Q(x,y)$
\be\label{Y1}
\tilde\psi_C(x,y)=\psi_Q(x)\psi^*_Q(y)=\rho_Q(x,y).
\ee
Reversing the partial Fourier transform \eqref{QC2} we obtain the classical wave function in phase space
\be\label{Y2}
\psi_C(z,p)=\int_re^{-ipr}\psi_Q\left(z+\frac r2\right)\psi^*_Q
\left(z-\frac r2\right)=\psi_C^*(z,p).
\ee
In consequence, the associated classical probability density $w=\psi^2_C$ obtains as 
\ba\label{Y3A}
&&w(z,p)=\int_{r,r'}
e^{ip(r'-r)}\\
&&\psi^*_Q\left(z+\frac{r'}{2}\right)
\psi_Q\left(z-\frac{r'}{2}\right)
\psi^*_Q\left(z-\frac r2\right)
\psi_Q\left(z+\frac r2\right).\nn
\ea
As it should be, this is a positive semidefinite function with proper normalization $\int_{z,p}w=1$. We recall that the classical probability to find a particle at point $z$
\ba\label{Y4}
w(z)=\int_pw(z,p)=\int_r
\left |\psi_Q\left(z+\frac r2\right)\right|^2
\left|\psi_Q\left(z-\frac r2\right)\right|^2\nn\\
\ea
differs from  the quantum probability $|\psi_Q(z)|^2$.

The association of a classical probability with a pure quantum state allows us to compute the expectation values of classical position and momentum observables in a pure quantum state
\ba\label{Y5A}
&&\kl F(X_{cl},P_{cl})\kr=\int_{z,p}F(z,p)w(z,p)\nn\\
&&=\int_{z,p,r,r'}F(z,p)
e^{ip(r-r')}\\
&&\psi^*_Q
\left(z+\frac{r'}{2}\right)\psi_Q
\left(z-\frac{r'}{2}\right)
\psi_Q^*\left(z-\frac r2\right)
\psi_Q\left(z+\frac r2\right).\nn
\ea
In contrast to the quantum observables, whose expectation values obtain  as quadratic expressions in $\psi_Q$, the expectation values of the classical observables involve four powers of the quantum wave function $\psi_Q(x)$ (here in position space). Nevertheless, they are well defined objects for any quantum state, such that we can compute a ``classical position'' and a ``classical momentum'' of the quantum particle, in addition to the quantum position and quantum momentum. The classical position and momentum are commuting observables - the order of $X_{cl}$ and $P_{cl}$ in the function $F(X_{cl},P_{cl})$ in eq. \eqref{Y5A} does not matter.

We recall that the classical wave function $\psi_C(z,p)$ equals the Wigner transform of the quantum mechanical density matrix, eq. \eqref{103B}. The classical observables can therefore also be written as
\be\label{Y7}
\kl F(X_{cl},P_{cl})\kr=\int_{z,p}F(z,p)\bar\rho^2_{w}(z,p).
\ee
Indeed, for a pure quantum state the Wigner transform of the density matrix has the properties required for a (classical) wave function 
\be\label{Y8}
\int_{z,p}\bar\rho^2_{w}(z,p)=1,
\ee
which also ensures that $\bar\rho^2_{w}$ is an acceptable classical probability distribution according to eq. \eqref{102C}. This contrasts with $\bar\rho_{w}$, which may be negative in certain regions of phase space. The expectation values of classical observables employ $\bar\rho^2_{w}$, which is a probability density, in contrast to the expectation values of totally symmetrized quantum observables which are linear in $\bar\rho_{w}$. For wave packets one may check explicitly that the pure state wave function $\psi_Q$ in eq. \eqref{189I}, inserted into eq. \eqref{Y2}, yields the classical probability distribution \eqref{189C} with $\Delta_x=1/(4\Delta_p)$, and that $w=\bar\rho^2_w$ with $\tilde \Delta_x=\sqrt{2}\Delta_x,\tilde\Delta_p=\sqrt{2}\Delta_p$ in eq. \eqref{189B}.

\medskip\noindent
{\bf 2. Mixed quantum states}

Our explicit construction shows that for every pure quantum state there exists a unique classical probability distribution such that the quantum particle can be realized by coarse graining. This can be generalized to mixed quantum states but the uniqueness will be lost. For an arbitrary density matrix in quantum  mechanics $\rho_Q(x,y)$ we can take the root in a matrix sense $\sigma(x,y)$ such that
\ba\label{Y9}
\rho_Q(x,y)&=&\int_v\sigma(x,v)\sigma(v,y),\nn\\
\sigma^*(x,y)&=&\sigma(y,x).
\ea
This can be seen easily by bringing $\rho_Q$ to diagonal form by a suitable unitary transformation. Positivity imples that all diagonal elements are positive semi-definite such that the root of the diagonal matrix is well defined. The relation $\rho_Q=\sigma^2$ is preserved by unitary transformations. We can now use $\sigma$ in order to define a classical wave function and density matrix
\ba\label{Y10}
\tilde\psi_C(x,y)&=&\sigma(x,y)~,~\psi_C(z,p)=\sigma(z,p)=\psi_C^*(z,p),\nn\\
w(z,p)&=&\sigma^2(z,p)~,~\rho_C(x,x',y,y')=\sigma(x,y)\sigma^*(x',y').\nn\\
\ea
It is easy to verify that insertion of this wave function into eq. \eqref{P4} yields indeed the relation \eqref{Y9}. This demonstrates that an arbitrary $\rho_Q$ can be obtained by coarse graining from the classical probability distribution $w$ in eq. \eqref{Y10}. 

We observe that a unitary transformation 
\ba\label{Y12}
\hat\psi_C(x,y)&=&(\tilde \psi_C U)(x,y)=\int_v\tilde\psi_C(x,v)U(v,y),\nn\\
&&\int_v U(x,v)U^*(y,v)=\delta(x-y),
\ea
does not affect the coarse grained density matrix
\ba\label{Y13}
\rho_Q(x,x')&=&\int_y\hat\psi(x,y)\hat\psi^*(x',y)\nn\\
&=&\int_{y,v,v'}\tilde\psi_C(x,v)U(v,y)U^*(v',y)\tilde\psi_C^*(x',v')\nn\\
&=&\int_v\tilde\psi_C(x,v)\tilde\psi_C^*(x',v).
\ea
We could therefore use $\hat \psi_C$ instead of $\tilde\psi$ for the definition of a classical wave function, such that the coarse graining of the classical density matrix still yields the same $\rho_Q$. However, for arbitrary $\hat\psi_C$ \eqref{Y12} the wave function $\hat\psi(z,p)$ is no longer real. Reality of $\psi_C(z,p)$ obtains from  the hermiticity property $\tilde\psi_C^*(x,y)=\tilde\psi_C(y,x)$, which holds for eq. \eqref{Y10}. 

The coarse graining \eqref{P4} involves an integration and therefore leads to a loss of information. We cannot reconstruct $\rho_C(x,x',y,y)$ from $\rho_Q(x,x')$ only using eq. \eqref{P4}. If we insist on the reality of the classical wave function $\psi(z,p)$, however, the possible choices of $\rho_C(x,x',y,y)$ get severely restricted. For example, all classical product states $\tilde\psi_C(x,y)=\psi_Q(x)\zeta(y)$ \eqref{T10} lead to the same quantum pure state with wave function $\psi_Q(x)$, while the hermiticity of $\tilde\psi_C(x,y)$ restricts $\zeta(y)=\psi^*_Q(y)$.

In addition to the reality of $\psi_C(z,p)$ we will also impose ``smoothness restrictions'' on $\psi_C(z,p)$ and $\tilde\psi_C(x,y)$ such that the expectation values of appropriate statistical observables are well defined by the quantum expressions of the type \eqref{113D}. For a given quantum state we would like to know what are the remaining possibilities for the quantum  transform \eqref{AA1} of the classical probability distribution $w(z,p)$ to yield the Wigner representation of a given quantum density matrix $\bar\rho_w(z,p)$. If for mixed quantum states there is more than one possibility for the choice of $w(z,p)$ which yields a given quantum  density matrix $\rho_Q$, the information contained in $\rho_Q$ will not be sufficient for a unique prediction of expectation values and correlations involving the classical position and momentum observables. Additional information which $w(z,p)$ describes the given mixed quantum state will be necessary for this purpose. 

We have already seen in a particular example that the map from the classical ensemble characterized by $w(z,p)$ to mixed quantum states is not injective - two different classical wave packets yield the same quantum wave packet in eqs. \eqref{189B}, \eqref{189C}, \eqref{189F}. This implies that, in general, the choice of classical observables is not unique for a quantum state. For a given mixed state density matrix $\bar\rho_w(z,p)$ one has to know to which classical state it is associated if one wants to compute the expectation values of the classical observables. In contrast, we have argued that for a pure quantum state a unique choice of classical observables is singled out by eq. \eqref{Y7}. In other words, for mixed states the information contained in the coarse grained $\bar\rho_w(z,p)$ is not sufficient in order to compute the expectation values of classical observables, while in the special case of pure quantum states the restriction of reality allowed classical wave functions is strong enough such that classical observables can be determined uniquely.

The issue may be understood best if one tries to construct a map from $\bar\rho_{w}(z,p)$ to a classical wave function $\psi_C(z,p)$. The map from $\rho_{w}(z,p)$ to $\rho_Q(x,y)$ and therefore to $\sigma^2(x,y)$ in eq. \eqref{Y9} is well defined. Ambiguities can arise, however, in the choice of $\sigma(x,y)$, while a given $\sigma(x,y)$ is again mapped uniquely to the classical wave function $\psi_C(z,p)$ by eq. \eqref{Y10}. In a basis where $\sigma^2$ is diagonal the elements of $\sigma$ are fixed only up to signs. The smoothness conditions will restrict the possible signs of these diagonal elements severely, but they are not always strong enough to admit only one unique ``smooth'' solution $\sigma(x,y)$. This explains why different classical ensembles may be mapped to the same quantum state by the quantum transform. An exception are pure quantum states. 

\section{Sharpened observables}

The discussion of the preceeding section raises a simple question: can classical observables be measured in a pure quantum state? From the point of view of the classical ensemble, described by the probability density $w(x,p)$, nothing speaks against this possibility. After a typical measurement of both the commuting observables $X_{cl}$ and $P_{cl}$ the ensemble changes to an ensemble with fixed $X_{cl}=\bar x,P_{cl}=\bar p$, and $\Delta^2_x\to 0,\Delta^2_p\to 0$ for the wave packet \eqref{189C}. After such a measurement, the state remains no longer a pure quantum state. Measurements of this type can perfectly be described by the quantum formalism for the classical particle, with classical wave function $\psi(z,p)$. Thus from the conceptual side there is no obstruction against the possibility to measure the classical observables. The issue reduces to the question if a measurement apparatus for the classical observables can be realized practically in an experiment.

This question can only be settled by observation. Assume one measures the probability of a particle to be at point $x$ for a given quantum state $\psi_Q(x)$ - a typical setting for a double slit experiment. The prediction for this measurement depends on the choice of the position observable. We may actually continuously interpolate between the quantum position $X_Q$ and the classical position $X_{cl}$ by 
\be\label{Y14}
X_\beta=\cos^2\beta X_Q+\sin^2\beta X_{cl}.
\ee
The probability to find the particle at location $x$ is given by
\ba\label{Y15}
p(x)&=&\int_{z,p}\psi(z,p)\delta(X_\beta-x)\psi(z,p)\nn\\
&=&\int_{z,p}\psi(z,p)\delta\big
((z+\frac i2\partial_p)\cos^2\beta+z\sin^2\beta -x\big)
\psi(z,p)\nn\\
&=&\int_{z,p}\psi(z,p)\delta(z+\frac i2\cos^2\beta\partial_p-x)\psi(z,p).
\ea
For the classical position $(\beta=\pi/2)$ one has \eqref{Y4}
\be\label{Y16}
p(x)=\int_pw(x,p)=\int_r\left|\psi_Q\left(x+\frac{r}{2}\right)\right|^2
\left |\psi_Q \left(x-\frac r2\right)\right|^2,
\ee
while for the quantum position $(\beta=0)$
\be\label{Y17}
p(x)=|\psi_Q(x)|^2=\int_p\bar\rho_{w}(x,p).
\ee
For arbitrary $\beta$ one finds a continuous interpolation
\be\label{Y18}
p(x)=\int_r\left|\psi_Q\left(x+\frac r2\sin^2\beta\right)\right|^2
\left|\psi_Q\left(x-\frac r2(1+\cos^2\beta)\right)\right|^2.
\ee
In a given experiment one should determine the parameter $\beta$ from the observed particle distribution $p(x)$. Precision tests of quantum mechanics may be conceived in this context as precision tests if the position is given by the quantum position observable. Their results can be quoted in terms of bounds on $\beta$. 

\section{Conclusions}
\label{Conclusions}

We have found a consistent description of a quantum particle in a potential which is based on the coarse graining of a ``classical'' probability distribution in phase space. For this purpose we propose a new fundamental dynamic equation which governs the time evolution of the probability density in phase space $w(z,p)$. This equation replaces the Liouville equation - the latter applies for a classical particle. The observables for position and momentum which are consistent with the coarse graining differ from the classical position and momentum observables. They involve statistical properties as the roughness of $w(z,p)$. 

We associate the coarse grained subsystem with the quantum probabilities for a particle in a potential, as expressed by the quantum density matrix $\rho_Q$. In turn, the observables consistent with the coarse graining can be associated with the quantum operators for position and momentum. They do not commute. Our modified basic dynamic equation for the classical probability density results on the coarse grained level in the von-Neumann equation for $\rho_Q$ which describes the time evolution for a quantum particle in a potential. 

We postulate that measurements of position and momentum are related to the coarse grained position and momentum observables, and that correlations of such measurements are predicted by the correlations of the associated quantum operators as appropriate for suitable measurement correlations based on conditional probabilities. Then all predictions for measurements of a quantum  particle in a potential can be expressed in terms of the classical probability distribution $w(z,p)$ and its time evolution. This includes all striking quantum effects as interference in a double slit experiment or tunneling through a potential barrier.

The conceptual and practical embedding of quantum physics in a probabilistic setting based on a classical statistical ensemble with positive probabilities in phase space permits us to address new types of questions. As one example, we have investigated the consequence of a different basic dynamical equation for $w(z,p)$ for the coarse grained subsystem. For this purpose we have studied the standard Liouville equation for $w$. On the coarse grained level this leads, in general, to a non-unitary time evolution of the quantum system which may describe decoherence or syncoherence. Under particular circumstances a unitary time evolution may be recovered, however with an effective Hamiltonian that differs from the standard quantum Hamiltonian for a particle in a potential. We have established the conditions for this to happen.

A second interesting issue concerns the choice of position and momentum observables that are associated to a given measurement procedure for a quantum system. We have found that a pure quantum state allows the definition of classical position and momentum observables in addition to the quantum observables. Also observables interpolating between the quantum and classical observables in dependence on a continuous parameter $\sin^2\beta$ can be defined consistently. The expectation values for such observables or the probability to find a particle with a given value of the interpolated position or momentum observable are computable in terms of the pure state quantum wave function. It becomes then an experimental question to find out which choice of observables corresponds to a given measurement device. In particular, precision tests of interference experiments of the general type of the double slit experiment can place bounds on $\sin^2\beta$ and determine quantitatively how well the association between measurements and quantum observables is realized.

\LARGE
\section*{APPENDIX A: DIFFERENT COARSE GRAININGS}
\renewcommand{\theequation}{A.\arabic{equation}}
\setcounter{equation}{0}

\normalsize
The coarse graining of the classical density matrix \eqref{P4} is not unique. Many different ways of taking a subtrace of a density matrix are possible. As an alternative we may not perform the Fourier transform \eqref{QC2} and define the coarse graining by a momentum trace over $\rho(z,z',p,p')$, 
\be\label{H1}
\tilde \rho(z,z')=\int_p\rho(z,z',p,p).
\ee
Performing a Fourier transform
\be\label{H2}
\rho(z,z',r,r')=\int_{p,p'}e^{ipr}e^{-ip'r'}
\rho(z,z',p,p')
\ee
we can equivalently perform the trace over $r$
\be\label{H3}
\tilde \rho(z,z')=\int_r\rho(z,z',r,r).
\ee
The coarse grained density matrix $\tilde \rho(z,z')$ defined in this way is different from the quantum density matrix $\rho_Q(x,x')$ defined by eq.\eqref{P4}. The Wigner transform of $\tilde \rho (z,z')$, i.e. 
\ba\label{HBA}
\tilde\rho_w(z,p)&=&\int d(z'-z'')e^{-ip(z'-z'')}\tilde\rho(z',z'')\nn\\
&=&\int_{z',z'',p'}\delta\left(\frac{z'+z''}{2}-z\right)\nn\\
&&e^{-ip(z'-z'')}\psi(z',p')\psi^*(z'',p'),
\ea
also differs from eq. \eqref{51}. In the $(z,r)$ basis the classical Hamiltonian reads
\be\label{H4}
H_L=-\frac{1}{m}\partial_z\partial_r+V'(z)r
\ee
and we obtain
\ba\label{H5a}
&&i\partial_t\tilde\rho(z,z')=-\frac 1m \int_{r,r'}\delta(r-r')
[\partial_z\partial_r-\partial_{z'}\partial_{r'}\nn\\
&&+\int_r\big(V'(z)-V'(z')\big)r]
\rho(z,z',r,r').
\ea
The discussion of the issue of unitary of this evolution can proceed along similar lines as for the coarse graining \eqref{P4}. 

Generalized classical particles, where $\rho(z,z',p,p')$ follows a time evolution different from the Liouville equation, seem particularly simple to formulate for the coarse graining \eqref{H1}: if $H_L$ is replaced by $H_Q(z)=-\partial^2_z/2m+V(z)$ the coarse grained density matrix obeys the von-Neumann equation with Hamiltonian $H_Q(z)$. However, a real $\psi(z,p)$ does not remain real under the action of $H_Q$, in contrast to the time evolution with $H_W$ (eq. \eqref{GP2}). A possible classical statistical interpretation of a complex wave function $\psi_C(z,p)$ in phase space, which would be needed for a description of such a time evolution remains to the investigated.

\section*{APPENDIX B: COMPLEX WAVE FUNCTION IN PHASE SPACE}
\renewcommand{\theequation}{B.\arabic{equation}}
\setcounter{equation}{0}
In this appendix we briefly investigate a setting with a complex classical wave function
\be\label{B.A}
\psi(z,p)=w^{1/2}(z,p)\exp\{i\alpha(z,p)\}.
\ee
The relation to the phase space probability involves in this case the absolute square
\be\label{B.B}
w(z,p)=|\psi(z,p)|^2.
\ee
The phase $\alpha$ influences the expectation values of the quantum and statistical observables, as
\be\label{QC22}
\kl P_Q\kr=\int_{p,z}w(z,p)(p+\frac 12\partial_z\alpha),
\ee
or
\ba\label{QC23}
\kl P^2_Q\kr&=&\int_{p,z}
\left[w(z,p)(p^2+p\partial_z\alpha)
+\frac14\partial_z\psi^*(z,p)\partial_z\psi(z,p)\right]\nn\\
&=&\int_{p,z}\left[w(z,p)\left(p+\frac12\partial_z\alpha\right)^2
+\frac{1}{16}
\frac{\big(\partial_zw(z,p)\big)^2}{w(z,p)}\right].\nn\\
\ea
For a complex classical wave function \eqref{B.A} we note the shift by $\partial_z\alpha/2$ as compared to the classical momentum.This may seem intriguing at first sight, since the classical probability distribution $w(z,p)$ does not contain any information about the phase $\alpha(z,p)$ in eq. \eqref{B.A}. A nonzero $\alpha$ corresponds in eq. \eqref{QC2} to a shifted Fourier transform if we express $\tilde\psi_C(x,y)$ in terms of $w^{1/2}(z,p)$. In turn, the meaning of ``integrating out the $y$-coordinate'' in the definition of the coarse graining \eqref{P4} depends on $\alpha$. Thus different choices of $\alpha$ define different coarse grainings of the classical probability distribution, explaining why expectation values of the quantum momentum can depend on $\alpha$.

A complex classical wave function would also affect the result of the quantum transform to a Wigner function.
\ba\label{QC27}
&&\bar\rho_w(z,p)=\int d(x-x')e^{-ip(x-x')}\bar\rho(x,x')\nn\\
&=&\int_{x,x',y}\delta\left(\frac{x+x'}{2}-z\right)
e^{-ip(x-x')}\psi(x,y)\psi^*(x',y)\nn\\
&=&\int_{x,x',y,q,q'}\delta\left(\frac{x+x'}{2}-z\right)\nn\\
&&\exp\Big\{i\big[q(x-y)-q'(x'-y)-p(x-x')\big]\Big\}\nn\\
&&\psi\left(\frac{x+y}{2},q\right)\psi^*
\left(\frac{x'+y}{2},q'\right)\nn\\
&=&\int_{r,r',s,s'}\sqrt{w(z+\frac r2~,~p+s)w\left(z+\frac{r'}{2},p+s'\right)}\nn\\
&&\cos \Bigg\{s'r-sr'+\alpha\left(z+\frac r2~,~p+s\right)\nn\\
&&~\qquad ~-\alpha\Big(z+\frac{r'}{2}~,~p+s'\Big)\Bigg\}.
\ea
Furthermore, if we generalize our setting to a complex wave function $\psi=\sqrt{w}\exp (i\alpha)$ the evolution equation also involves $\alpha$,
\ba\label{GP4}
\partial_tw&=&-\hat L w-\frac\lambda8 z
\Bigg\{\frac34 w^{-2}(\partial_p w)^3-\frac32 w^{-1}
\partial_p w\partial^2_pw+\partial^3_pw\nn\\
&&-3\partial_pw\partial^2_p\alpha
 -6w\partial_p\alpha\partial^2_p\alpha\Bigg\}.
\ea
While a constant phase $\alpha$ remains irrelevant in all these relations, a momentum or position dependent $\alpha$ can store additional information and influences the dynamics and expectation values by modifications $\sim \partial_z\alpha$ or $\partial_p\alpha$.


\end{document}